\documentclass[tighten,iop]{emulateapj}

\usepackage[english]{babel}
\usepackage{amsmath}
\usepackage{amsfonts}
\usepackage{amssymb}
\usepackage{graphicx}
\usepackage[dvips]{color} 

\newcommand{\nuc}[2]{${}^{#2} \rm #1$}

\def\gtaprx {\lower .14ex\hbox{\rlap{\raise .9ex\hbox{\hskip .3ex
	{\ifmmode{\scriptscriptstyle >}\else
		{$\scriptscriptstyle >$}\fi}}}
	\kern -.4ex{\ifmmode{\scriptscriptstyle \sim}\else
		{$\scriptscriptstyle\sim$}\fi}}}
\def\ltaprx {\lower .14ex\hbox{\rlap{\raise .9ex\hbox{\hskip .3ex
	{\ifmmode{\scriptscriptstyle <}\else
		{$\scriptscriptstyle <$}\fi}}}
	\kern -.4ex{\ifmmode{\scriptscriptstyle \sim}\else
		{$\scriptscriptstyle\sim$}\fi}}}

\newcommand{\s}{ \, {\rm s} }
\newcommand{\K}{ {\,\rm K} }

\newcommand{\km}{{\, \rm km}}

\newcommand{\del}[2]%
{\frac{\mathrm{d}{#2}}{\mathrm{d}{#1}}}
\newcommand{\Del}[2]%
{\frac{D{#2}}{D{#1}}}\newcommand{\ddel}[2]%
{\frac{\mathrm{d}^2{#2}}{\mathrm{d}{#1}^2}}

\newcommand{\pdel}[2]%
{\frac{\partial{#2}}{\partial{#1}}}
\newcommand{\pddel}[2]%
{\frac{\partial^2{#2}}{\partial{#1}^2}}

\newcommand{\laplace}{\bigtriangleup}

\renewcommand{\vec}[1]{\mathbf{#1}}

\shorttitle{Post-shock-revival evolutions of core-collapse supernovae}
\shortauthors{Yamamoto et al.}

\begin{document}

\title{Post-shock-revival evolutions in the neutrino-heating mechanism of
core-collapse supernovae}

\author{
Yu Yamamoto\altaffilmark{1},
Shin-ichiro Fujimoto\altaffilmark{2},
Hiroki Nagakura\altaffilmark{3,4}
and
Shoichi Yamada\altaffilmark{1,4}
}

\altaffiltext{1}{Science \& Engineering, Waseda University,
3-4-1 Okubo, Shinjuku, Tokyo 169-8555, Japan}

\altaffiltext{2}{
 Kumamoto National College of Technology, 
 2659-2 Suya, Goshi, Kumamoto 861-1102, Japan
}

\altaffiltext{3}{Yukawa Institute for Theoretical Physics, Kyoto
  University, Oiwake-cho, Kitashirakawa, Sakyo-ku, Kyoto, 606-8502,
  Japan
}

\altaffiltext{4}{Advanced Research Institute for Science \&
Engineering, Waseda University, 3-4-1 Okubo,
Shinjuku, Tokyo 169-8555, Japan}


\begin{abstract}
We perform some experimental simulations in spherical symmetry and 
axisymmetry to understand the post-shock-revival evolution of core-collapse
supernovae. Assuming that the stalled shock wave is relaunched by neutrino
heating and employing the so-called light bulb approximation, we induce
shock revival by raising the neutrino luminosity by hand up to the critical value,
which is also determined by dynamical simulations. 
A 15${\rm M}_{\odot}$ progenitor model is employed. 
We incorporate
nuclear network calculations with a consistent equation of state in the 
simulations to account for the energy release by nuclear reactions and their feedback 
to hydrodynamics. Varying the shock-relaunch time rather arbitrarily, we
investigate the ensuing long-term evolutions systematically, paying particular 
attention to the explosion energy and nucleosynthetic yields as a function
of this relaunch time, or equivalently the accretion rate at shock revival. 
We study in detail how the diagnostic explosion energy approaches the asymptotic value 
and which physical processes contribute to the explosion energy 
in what proportions as well as their dependence on the relaunch time and the 
dimension of dynamics. We find that the contribution of nuclear reactions to the
explosion energy is comparable to or greater than that of neutrino heating. 
In particular, recombinations are dominant over burnings in the contributions of 
nuclear reactions. Interestingly 1D models studied in this paper cannot produce 
the appropriate explosion energy and nickel mass simultaneously, overproducing nickels,
whereas this problem is resolved in 2D models 
if the shock is relaunched at 300-400ms after bounce.
\end{abstract}





\section{Introduction}

The mechanism of core-collapse supernovae (CCSNe) has refused our understanding for the last half century. 
CCSNe are initiated as the implosion of central cores of massive stars at the end of their lives. The inward motion
is halted when the density exceeds the nuclear saturation density, $\sim 3 \times 10^{14}$g/cm$^3$, above which 
matter becomes very stiff with the adiabatic index of $\Gamma \gtrsim 2$. At this point, the core is still lepton-rich 
with the lepton fraction of $Y_{\ell}\sim 0.3$ and very hot with the temperature of a few tens MeV compared with 
the ordinary neutron star, which is the end product of CCSNe. The gravitational energy liberated so far is 
$\sim 10^{53}$erg, far greater than the typical kinetic energy of ejecta in supernova explosions, $\sim 10^{51}$erg. 
This large energy is mostly stored in the core as internal energy and unavailable for the prompt explosion initiated 
by the shock wave produced at the core bounce. As a result, the shock wave is soon stagnated in the core owing to 
dissociations of nuclei and neutrino cooling. The current study on the mechanism of CCSNe is hence focused on the revival
of the stalled shock wave.

There are a couple of viable mechanisms proposed at present~\citep{2012arXiv1204.2330K,2012arXiv1206.2503J}. The most promising among them is 
supposed to be the neutrino heating mechanism. The large energy reservoir in the core is tapped by neutrinos over the next ten seconds. 
In this scenario, a part of the energy carried away by neutrinos are deposited in the so-called heating region, where
the heating of matter via absorptions of neutrinos emitted by the proto neutron star overwhelms the cooling through 
local emissions of neutrinos. If the energy deposition is large enough, the accretion shock wave will be reinvigorated and
restart to propagate outward, expelling the stellar envelope and culminating in well-known optical displays of supernovae 
when the shock wave reaches the stellar surface. In fact, it is now almost established that there is a critical luminosity 
for the revival of the stalled shock wave, which is a function of the rate of matter accretion: the more rapidly matter  
accretes, the greater the luminosity should be for the shock revival~\citep{1993ApJ...416L..75B,1996ApJ...473L.111K,2005ApJ...623.1000Y,2008ApJ...688.1159M,2010ApJ...720..694N,2012ApJ...755..138H}.

\citet{1993ApJ...416L..75B} was the first to point out the existence of the critical luminosity, calculating a sequence of 
spherically symmetric, steady, shocked accretion flows and showing the non-existence of such flows for luminosities beyond
a certain threshold given for each mass accretion rate. \citet{2005ApJ...623.1000Y} analyzed the sequence more in detail and showed 
that there are in general two accretion flows for a given accretion rate, with one of them with a larger shock radius being always unstable to 
radial perturbations and the other being stable and that the branches join with each other at the critical luminosity~\citep{2012ApJ...746..106P}. 
Recently \citet{2012ApJ...746..106P} found that the so-called antesonic condition predicts accurately the point, at which the spherically symmetric, steady,
shocked accretion flow ceases to exist. \citet{2006ApJ...641.1018O} demonstrated for the spherical case that the accretion shock may be 
revived by overstabilizing radial oscillations before the point of the non-existence of steady shocked flows is reached, 
the fact also confirmed recently by \citet{2012ApJ...749..142F}. 

For more realistic, non-spherical flows, the critical luminosity is lowered~\citep{1996astro.ph.10214.,2006ApJ...641.1018O,2008ApJ...688.1159M,2010ApJ...720..694N,2012ApJ...755..138H}.
Non-radial instabilities referred to as the standing accretion shock instabilities or SASI are known to set in for lower neutrino 
luminosities than the radial instabilities~\citep{2001A&A...368..311F,2005ApJ...623.1000Y}. They effectively enhance the neutrino heating by dredging 
up heated material and channeling cold material. It has been demonstrated that the ratio of the residence time to the heating time
is a good measure to gauge how close to shock revival a particular accretion flow is~\citep{2001A&A...368..527J,2003ApJ...592..434T,2008ApJ...688.1159M}. Here the
residence time is a duration, in which each mass element lingers in the heating region and the heating time is a time needed to 
deposit enough energy to unbound the mass element. This criterion is rather easy to evaluate and has been applied particularly for
numerical results~\citep{2009ApJ...694..664M,2008ApJ...688.1159M,2010ApJ...720..694N,2012ApJ...755..138H,2012ApJ...749...98T}.

It should be emphasized that the revival of stalled shock waves is not sufficient to ascertain the supernova mechanism. 
Although there are some groups reporting the canonical explosions
~\citep{bruenn2013}, 
almost all numerical simulations that have reported successful shock revival have yielded energies of ejecta that are substantially 
smaller than the canonical value of $10^{51}$erg 
~\citep{muller2012,suwa2013}. 
Although the authors claim that the energies are still increasing at the end of
their simulations and might reach appropriate values should the computations be continued long enough, a convincing demonstration
remains to be done. In addition to the explosion energy, the mass of neutron star as well as nucleosynthetic yields should be also
reproduced properly in the successful supernova simulations. 

In this paper, we are interested in what will happen after the stagnated shock wave is successfully relaunched by the neutrino 
heating. In particular we discuss (1) when the explosion energy is determined, (2) which processes contribute to the explosion
in what proportions, (3) how the explosion energy and neutron star mass are dependent on the timing of shock revival, and (4) how
multi-dimensionality affects all of these issues. For these purposes, we have done a couple of numerical experiments in 1D (spherical
symmetry) and 2D (axisymmetry). Controlling neutrino luminosities under the light bulb approximation, we have induced shock revival
from different points on the critical curve (the critical luminosity as a function of the mass accretion rate) and computed the
following evolutions of matter flows outside the proto neutron star long enough for the energy of ejecta to become constant. 
In so doing we have taken into account nuclear reactions in a manner consistent with the EOS employed. The feedback from the reactions 
to hydrodynamics is fully incorporated. These consistencies were lacked in previous works~\citep{2009ApJ...694..664M,2009ApJ...703.1464F,2012arXiv1205.3657U,2012arXiv1207.5955N}.

The paper is organized as follows. In the next section, we describe the models and numerical details together with the assumptions
and approximations employed in this paper. The main results are presented in Section~\ref{sec:results} first for the spherically symmetric 
1D case and then for the axisymmetric 2D case. The summary and some discussions are given in the last section.

\section{Setup}
\label{sec:setup}

\subsection{Outline}
Before going into the details of our model building, we give a brief description of what we are going to do,
emphasizing the underlying ideas. 

We are interested in what happens after the relaunch of the stalled shock wave. The investigations in this paper 
are of experimental nature. We assume that the neutrino heating mechanism works successfully, which implies 
that the neutrino luminosity and accretion rate should be located on the critical curve just at the
shock revival. Exactly at which point on the curve the shock is relaunched is still uncertain as mentioned 
earlier, however. We hence take the neutrino luminosity (or equivalently the accretion rate) at the shock 
revival as a free parameter and vary it arbitrarily to see how the ensuing physical processes are affected. 
We prepare a couple of initial conditions, which correspond to the points for different neutrino luminosities
on the critical curve. We then solve the hydrodynamical equations together with nuclear reactions in 1D 
(spherical symmetry) and 2D (axial symmetry) to obtain the ensuing evolutions. See
Section~\ref{sec:step2} for more 
details on how to trigger the shock revival.

We do not solve the evolution of the central high-density region, in which a proto-neutron star sits, but 
replace it by appropriate inner boundary conditions. The temporal variation of mass accretion rate is obtained 
by the computation of the infall of a realistic stellar envelope after the loss of pressure support at
the inner boundary and is employed for the preparation of the initial states and subsequent hydrodynamical 
simulations. This also enables us to use the mass accretion rate as a clock. We follow the post-revival evolutions
long enough so that the so-called diagnostic explosion energy is settled to the terminal value. Integrating 
the heating rates both by neutrino absorptions and nuclear reactions, we also obtain each contribution to 
the explosion energy. The nuclear reactions are divided into the recombinations of free nucleons to heavy nuclei 
 and the explosive nuclear burnings and we estimate their contributions separately. We further distinguish the 
recombinations that occur in the nuclear statistical equilibrium (NSE) and those out of equilibrium. By so doing,
we can pin down what contributes to the explosion energy in what proportions. We can also find the dependence of
the results on the timing of shock relaunch as well as the dimensionality of dynamics. 

In the following we give the details of our modeling. The basic equations, input physics and numerical methods
are described first. Then the preparation of the initial conditions, which requires multiple steps to avoid full
computations of the collapse to bounce to shock stagnations, will then be presented in detail.  
\subsection{Basic equations, input physics \& numerical methods\label{sec:method}}
Throughout this paper we neglect relativity and employ Newtonian equations of motion.
Not only for the post-relaunch evolutions but also for the preparations of the initial 
conditions we solve the following equations:
\begin{equation}
\Del{t}{\rho}+\rho\nabla\cdot\vec{v}=0,
\label{eq:cont}
\end{equation}
\begin{equation}
\rho \Del{t}{\vec{v}}=-\nabla P -\rho \nabla (\Phi +\Phi_{c}),
\label{eq:eom}
\end{equation}
\begin{equation}
 \rho \frac{D}{Dt}\displaystyle{\Bigl( \frac{e}{\rho} \Bigr)}
  = - P \nabla \cdot \vec{v} + Q_{\nu}, 
  \label{eq:energy}
\end{equation}
\begin{equation}
 \Del{t}{Y_e} = \frac{N_{\nu}}{\rho N_A},
  \label{eq:ye_flow}
\end{equation}
\begin{equation}
 \Del{t}{Y_i} = f_i\,(\rho, e, \{Y_{i}\}),
  \label{eq:yi_flow}
\end{equation}
where $\rho$, $P$, $\vec{v}$, $e$, $Y_e$, $Y_i$ and $N_A$ are mass density, pressure
fluid velocity, energy density, electron fraction, number fraction of nucleus $i$ and 
Avogadro's number, respectively. We denote the Lagrange derivative as $D/Dt$. 
Note that the energy density in Eq.~(\ref{eq:energy}) includes the rest mass energy and 
the energy production by nuclear reactions are thus taken account.

In Eq.~(\ref{eq:eom}), it is expressed explicitly that the gravitational potential has two
contributions, $\Phi$ from the accreting matter and $\Phi_{c}$ from a central object, whose 
mass, $M_{\rm in}$, is a function of time and calculated by the integration of mass 
accretion rates at the inner boundary of computational domain. They satisfy the 
following equations: 
\begin{equation}
\laplace{\Phi} = 4\pi G \rho,
\end{equation}
and
\begin{equation}
 \Phi_c = - \frac{G M_{\rm in}}{r},
\end{equation}
where $G$ is the gravitational constant.

$Q_{\nu}$ and $N_{\nu}$ in Eqs.~(\ref{eq:energy}) and (\ref{eq:ye_flow}) are
the source terms that give the rates of the changes in specific energy density and electron fraction, 
respectively, owing to weak interactions. In the present study, we take into account only absorptions and 
emissions of electron-type neutrinos and anti-neutrinos on nucleons. The weak interaction rates are adopted 
from \citet{2006A&A...457..963S}. We employ in this paper the so-called light bulb approximation, in which the computation 
of neutrino transport is neglected and neutrinos are just assumed to be emitted isotropically from the 
neutrino sphere with a given luminosity (Eq.~(\ref{eq:lum})) and energy spectrum, which we assume to have the Fermi-Dirac 
distribution~\citep{2006ApJ...641.1018O}. The radius of neutrino sphere is a function of time and is assumed 
in this paper to be given as follows~\citep{2006A&A...457..963S}:
\begin{equation}
R_{\nu}(t) = \frac{R_{\nu, i}}{1 + (1 - \exp (-t/t_{c}))(R_{\nu,i}/R_{\nu,f} - 1)},
\label{eq:nusp}
\end{equation}
where $R_{\nu, i}$ and $R_{\nu, f}$ are the initial and final values, respectively, and $t_{c}$ is 
the characteristic time scale. They are set to be $R_{\nu,i} = 58{\rm km}$ for $\nu_e$, $52{\rm km}$ for $\bar{\nu}_e$,
$R_{\nu, f} = 15{\rm km}$ and $t_c = 800{\rm ms}$. The temperatures, $T_\nu$, in the Fermi-Dirac distributions for the electron-type
neutrino and anti-neutrino are chosen so that their average energies would be 
$\langle \varepsilon_{\nu_e} \rangle = 20-8.0\times (1/2)^{t/200{\rm ms}}{\rm MeV}$
and $\langle \varepsilon_{\bar{\nu}_e} \rangle = 23-8.0\times (1/2)^{t/200{\rm ms}}{\rm MeV}$~\citep{2005ApJ...629..922S}.
The chemical potentials are assumed to be zero. In evaluating the heating and cooling of matter by neutrino absorptions and emissions, 
we employ the local distribution function of neutrino given by the following expression:
\begin{equation}
f(r, \varepsilon) = \frac{C(r)}{1+\exp(\epsilon /k_BT_\nu)},
\label{eq:dist}
\end{equation}
where $k_B$ is the Boltzmann's constant and the normalization factor, $C(r)$, is determined so that the local number density of neutrino,
$n_\nu(r)$, is given by the following relation:
\begin{equation}
L_{\nu} = 4\pi r^{2}\,\cdot c \cdot n_\nu(r) \cdot \langle \varepsilon_{\nu} \rangle \cdot \langle \mu (r)\rangle ,
\label{eq:lum}
\end{equation}
where $L_{\nu}=L_{\nu _{e}}=L_{\bar{\nu} _{e}}$ is the neutrino 
luminosity and 
the last factor, $\langle \mu (r) \rangle$, is the flux factor that accounts for how quickly the angular distribution 
becomes forward-peaked. We again employ the fitting formula given in \citet{2006A&A...457..963S} for the radial dependence of the flux factor. 

Eq.~(\ref{eq:yi_flow}) expresses the nuclear reactions. 

We deploy 28 nuclei: n, p, D, T, $^{3}$He, $^{4}$He 
and 12 $\alpha$-nuclei, i.e., $^{12}$C, $^{16}$O, $^{20}$Ne, $^{24}$Mg, 
$^{28}$Si, $^{32}$S, $^{36}$Ar, $^{40}$Ca, $^{44}$Ti, $^{48}$Cr, 
$^{52}$Fe, $^{56}$Ni, and 10 their neutron-rich neighbors, 
that is \nuc{Al}{27}, \nuc{P}{31}, \nuc{Cl}{35}, \nuc{ K}{39}, 
\nuc{Sc}{43}, \nuc{ V}{47}, \nuc{Mn}{51}, \nuc{Fe}{53}, 
\nuc{Fe}{54} and \nuc{Co}{55}.
We take into account emissions of a nucleon and $\alpha$ particle as one-body interactions as well as 
three-body reactions such as $3\alpha \rightarrow {\rm C}$ in addition to the main reactions: $(\alpha, \gamma)$, $(\alpha, p)$, 
$(p, \gamma)$ and their inverses. The reaction rates are taken from REACLIB~\citep{2000ADNDT..75....1R}. As demonstrated later, the employment 
of this rather small network is validated by the re-computations of nuclear yields with a larger network including 
463 nuclei, from n, p, and \nuc{He}{4} up to \nuc{Kr}{94}~\citep{fujimoto04}
for the densities and temperatures obtained by the simulations as a post-process. We combine the semi-implicit scheme developed by 
\citet{1999ApJS..124..241T} and the full implicit method to numerically handle Eq.~(\ref{eq:yi_flow}). The network calculations are needed 
only at low temperatures, since the
NSE
 is achieved at high temperatures and the compositions 
are determined so that the free energy should be minimized. In fact, the network computations are expensive when the NSE is 
established. We hence solve the network only for $T<7\times10^9{\rm K}$ and otherwise calculate the NSE compositions for the 
same 28 nuclei as used for the network computations. This temperature is high enough to ensure the establishment of NSE. 

The equation of state we use in this paper is the sum of the contributions from nucleons, nuclei, photons, electrons and positrons. 
The first two are treated as ideal Boltzmann gases, the composition of which is obtained either by the network computations or by the NSE
calculations as mentioned above. The photons are an ideal Bose gas and easy to handle whereas the electrons and positrons are 
treated as ideal Fermi gases, in which arbitrary degeneracy and relativistic kinematics are fully taken into account~\citep{1996ApJS..106..171B}.
It should be repeated that we include the rest mass contribution in the energy density so that the energy release by nuclear reactions
is automatically taken into account properly.

\begin{figure}[ht]
\epsscale{1.3}
 \plotone
{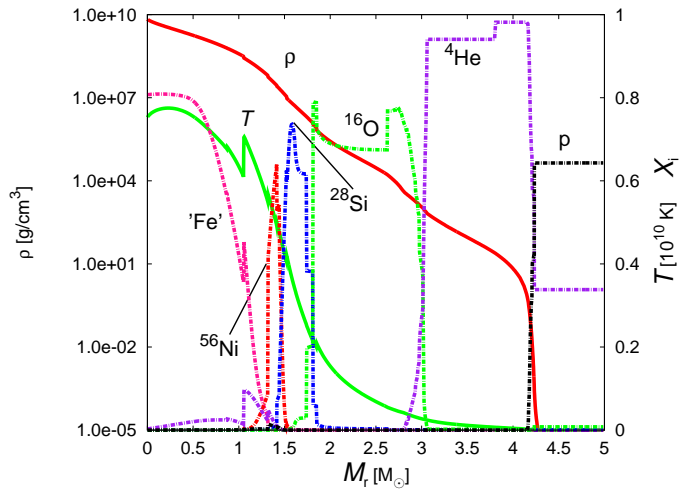}
\caption{The initial profile of the $15{\rm M}_{\odot}$ progenitor with 
the iron core of ${\rm M}_{\rm Fe}=1.4{\rm M}_{\odot}$. 
The density (red solid line) and temperature (green solid line) as well as mass fractions of 
some representative nuclei (solid dotted lines) are shown.}
\label{fig:fig1}
\epsscale{1.3}
\end{figure}

The hydrodynamical evolutions are solved with the ZEUS-2D code~\citep{1992ApJS...80..791S,2006ApJ...641.1018O} on the spherical coordinates. For the 1D 
computations, we use the same code just suppressing motions in the $\theta$ direction. We adopt a $15{\rm M}_{\odot}$ progenitor 
model computed by \citet{2007PhR...442..269W}. The profile just prior to collapse is shown in Fig.~\ref{fig:fig1}.
 In this paper we concentrate on this canonical model. The dependence of the results on the 
progenitor structures and the details of our modeling will be reported elsewhere. Note that according to the recent 
observations~\citep{2009MNRAS.395.1409S,2009ARA&A..47...63S,2011MNRAS.412.1522S} of core-collapse supernovae as well as their progenitors, the $15{\rm M}_{\odot}$ star may be
a typical progenitor of Type-II supernovae.


\subsection{Step 1: 1D simulation of the infall of envelope\label{sec:step1}}
We now proceed to the description of the first step in the preparation of initial models.
The aim of this step is to sample the mass accretion rates as a function of time together 
with the changes in the structure and composition of the envelope. For this purpose we 
perform a 1D simulation of the spherically symmetric implosion of the stellar 
envelope. We excise the interior of $r=60{\rm km}$ and replace it with the inner boundary,
at which we impose the free inflow condition. Since no bounce occurs in this simulation,
no shock wave emerges. Note that what we need is the accretion rate and the structure outside
the stalled shock wave, which would be produced and stalled somewhere inside the core in reality
and that they are unaffected by what happens inside the shock wave, since they are causally 
disconnected. The location of the inner boundary is chosen so that they would always
reside inside the shock wave.

We deploy 500 grid points to cover the region extending up to $r=2\times10^{5}{\rm km}$.
This is large enough to ensure that matter outside the outer boundary does not move essentially
for $\sim$ a second during this stage. 
The weak interactions are turned off for this computation, since the they are indeed negligible 
in the infalling envelope. Note again that the computational results for the region that 
would be engulfed by the shock wave in reality are irrelevant and do not have any consequence 
on the results outside. Hence the neglect of neutrino heating and cooling is completely justified. 
The nuclear reactions for the 28 nuclei, on the other hand, are computed for the region with
$T<7\times 10^9{\rm K}$ to follow the change in chemical composition and take account of its 
influence on the hydrodynamics during the implosion. The NSE composition is calculated for higher
temperatures.

\begin{figure}[ht]
\epsscale{1.2}
 \plotone
{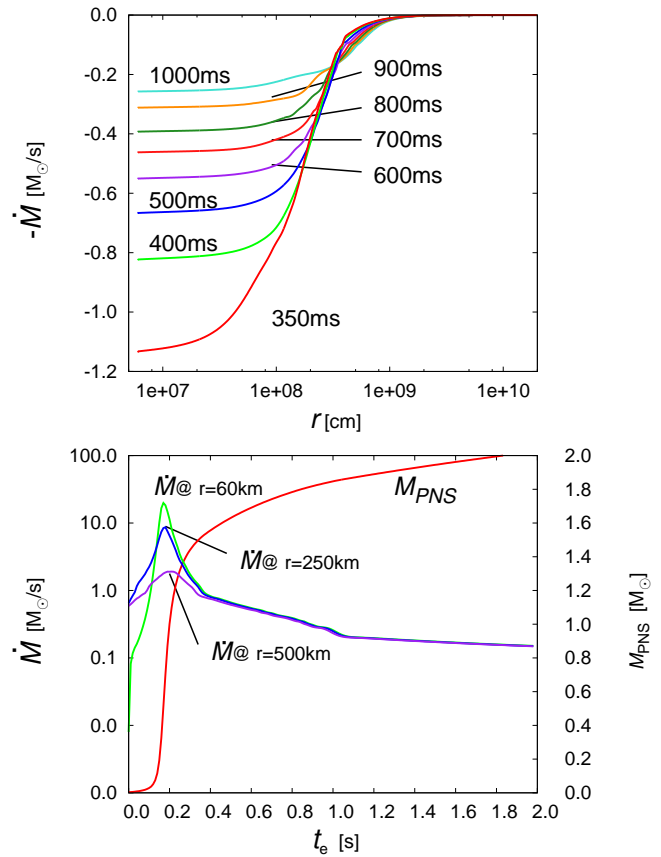}
\caption{The characters of the mass accretion rates.
The upper panel shows the radial profiles of $\dot{M}$ for different elapsed times, $t_e$. 
The lower panel displays the time evolutions of $\dot{M}$ at 3 different radii 
together with the proto-neutron star mass, $M_{PNS}$.}
\label{fig:fig2}
\epsscale{1.2}
\end{figure}

We show the results in Fig.~\ref{fig:fig2}. In the upper panel we show the mass accretion 
rate, $\dot{M} = 4\pi r^2 \rho v_r$, as a function of radius for different
$t_e$, which is the elapsed time from the beginning of collapse.
It is seen that the rarefaction wave generated by the inflow at the inner boundary propagates 
outward, triggering the infall of matter at large radii. After $t_e \sim 300{\rm ms}$ the accretion 
rates at $r \lesssim 500{\rm km}$ become independent of radius. This implies that the flows in 
this region can be approximated by steady accretions. The region is actually expanded outward
gradually. The lower panel shows the accretion rates at three different radii as a function
of time. As pointed out right now, they coincide with each other after $t_e \sim 300{\rm ms}$.
Before this time, on the other hand, the accretion rate is higher at smaller radii. There appears 
a peak at
 $t_e \sim 180{\rm ms}$
, which is rather insensitive to the radius. From a comparison with
realistic simulations~\citep{2007PhR...442...38J}, we find that this time roughly corresponds to the core bounce.
We hence refer to as the time elapsed from this point as the post-bounce time hereafter,
i.e, $t_{pb} \equiv t_e - t_{e(p)}$, where $t_{e(p)}$ denotes as the time of peak accretion rate. In the
same panel we also show the mass that has flown into the inner boundary by the given time, to which
we refer as the proto-neutron star (PNS) mass ($M_{PNS}$). In the bottom panel we present the time evolution in the plain of mass accretion rate and PNS mass.

\begin{figure*}
\vspace{15mm}
\epsscale{1.0}
\plottwo{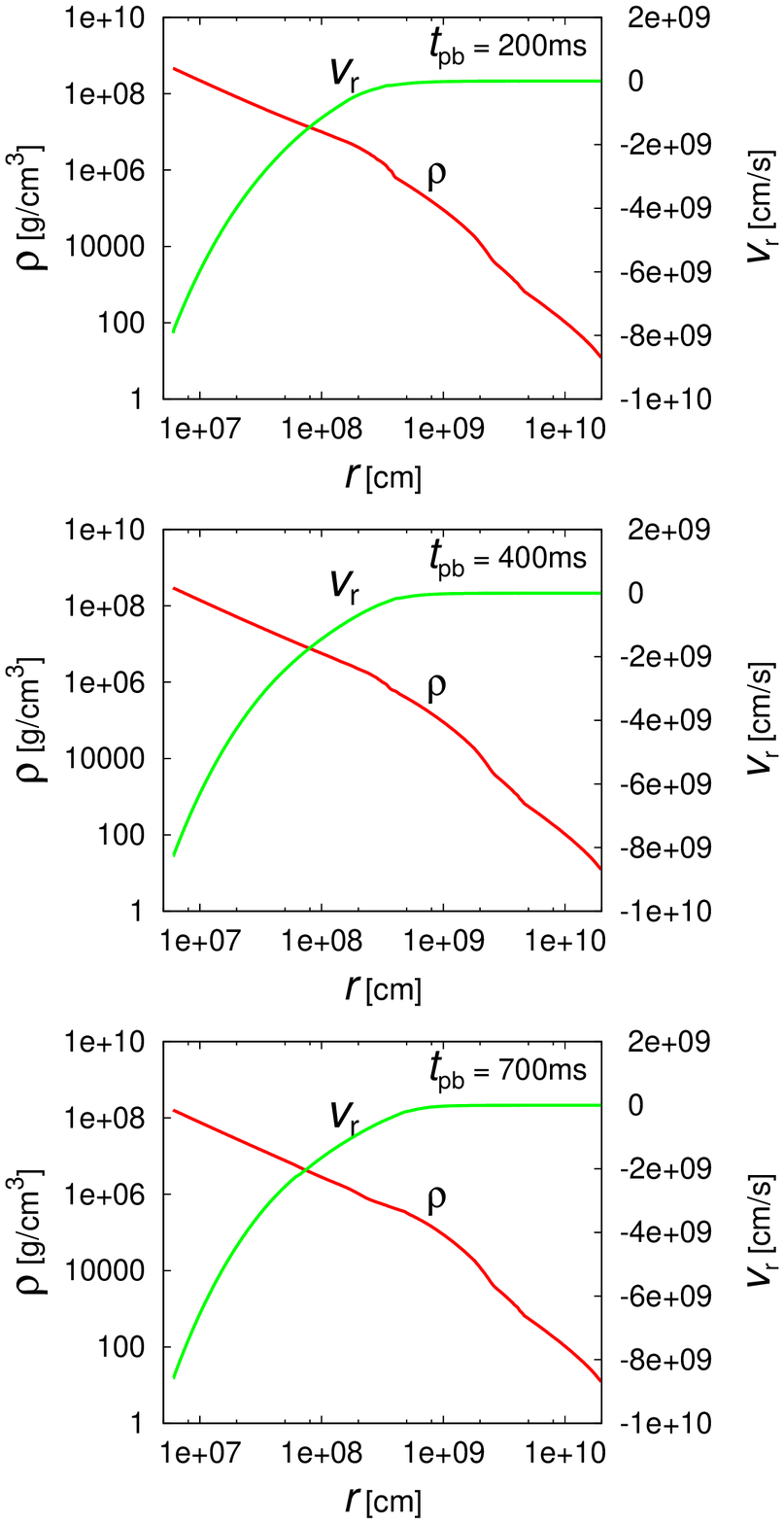}{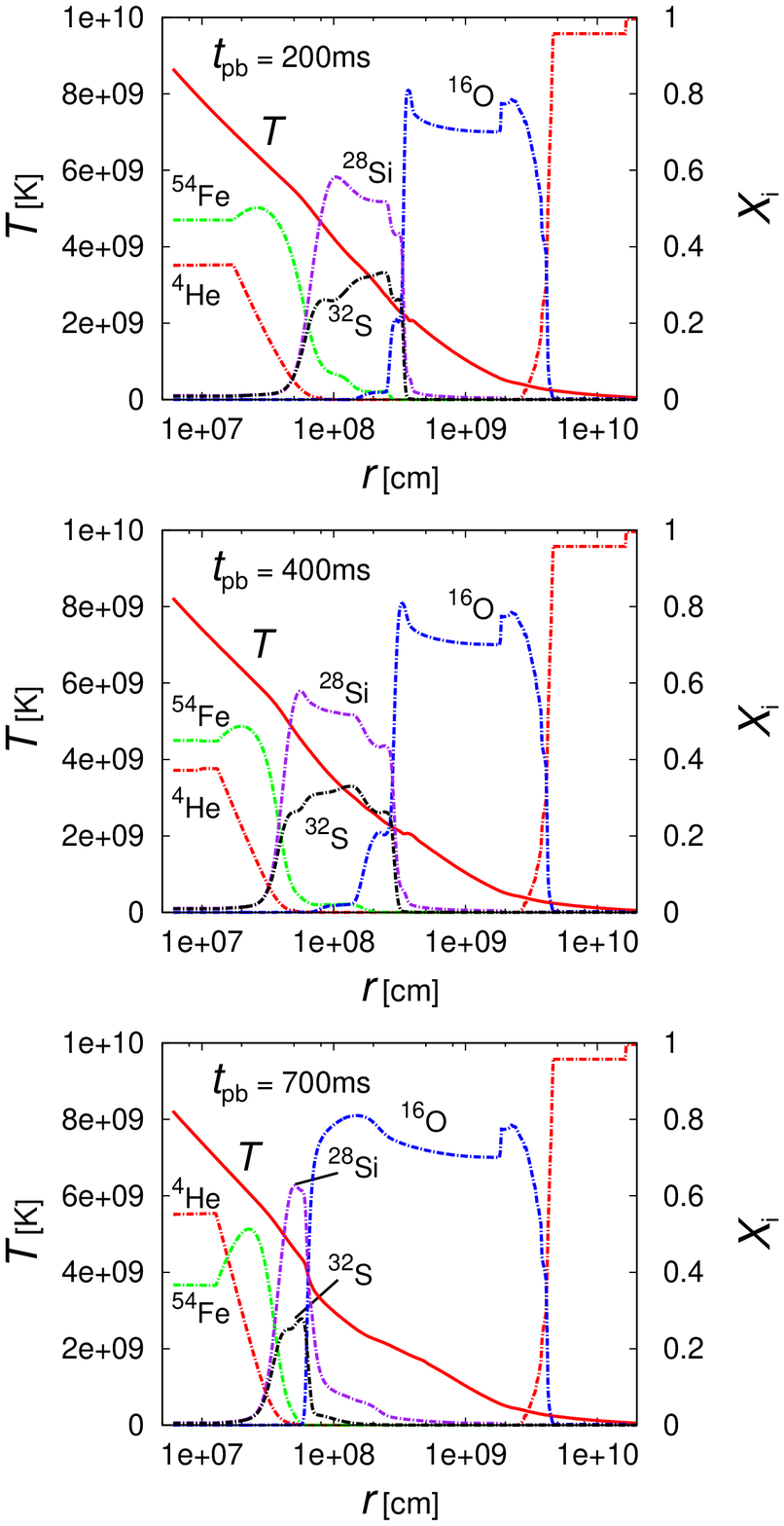}
\caption{The evolutions of the chemical composition as well as density, velocity and temperature.
In the left column the density (red line) and velocity (green line) are displayed for three different times, $t_{pb} =$ 200, 400 and 700ms, respectively. 
In the right column, the abundances of some representative nuclei (solid dotted lines) are shown with the temperature (solid line) 
for the same three times.
\label{fig:fig3}}
\epsscale{1.0}
\end{figure*}

The evolutions of the chemical composition together with the density, velocity and temperature are 
displayed in Fig.~\ref{fig:fig3}. In the left column the density and velocity are shown for three
different post-bounce times ($t_{pb}$). As the time passes, the density at a fixed radius decreases monotonically
whereas the inflow velocity gets larger. The outward propagation of the rarefaction wave is also
recognized in the figure, which is exactly how the implosion of envelope proceeds. In the right 
column, the chemical composition and temperature are presented for the same
$t_{pb}$.
The temperature at a fixed radius is in general a decreasing function of time. It is observed that 
heavy elements are advected inward. In addition, the changes in composition by nuclear burnings 
are also taken into account in this figure. 

We employ these results not only at the next step in the preparation of initial conditions, which
we will describe in the next section, but also for the simulations of the post-relaunch evolutions,
the results of which will be presented in \S~\ref{sec:results}.
 
\subsection{Step 2: search of critical luminosities\label{sec:step2}}
The aim of this step is to construct the critical steady accretion flows with a standing shock wave 
for the mass accretion rates obtained in step~1. It is important here to define unambiguously 
the critical point for a given accretion rate, since in this paper it is not meant for the flow
with the luminosity, above which no steady accretion flow exists~\cite{1993ApJ...416L..75B}. Instead we define
it to be the flow, in which the stalled shock wave is actually relaunched within a certain time.
This is because the shock revival normally occurs owing to hydrodynamical instabilities even in 1D 
before the luminosity, above which no steady accretion is possible, is reached. Hence we determine
the critical point hydrodynamically by following the growths of the instabilities for initially 
spherically symmetric and steady accretion flows.

We hence adopt a two-step procedure. In the first step, we construct a sequence of spherically 
symmetric and steady accretion flows with a standing shock wave for a given mass accretion rate.
We solve Eqs.~(\ref{eq:cont})-(\ref{eq:yi_flow}) in 1D, dropping the Eulerian time derivatives.
At the shock wave, we impose the Rankine-Hugoniot jump conditions. The nuclear reactions and weak
interactions that are described in \S\ref{sec:method} are fully taken into account. The outer
boundary of the computational domain is set to $500{\rm km}$ and the values of various quantities
are taken from the results of Step~1 at the times that correspond to the given mass accretion 
rates (see the bottom panel of Fig.~\ref{fig:fig2}). At the inner boundary, which is placed 
according to Eq.~(\ref{eq:nusp}) with the post-bounce time obtained from the mass accretion rate,
we impose the condition that the density be $\rho=10^{11}{\rm g/cm}^{3}$.
We cover the 
computational region with $300$ grid points. As the neutrino luminosity is increased, the location
of the standing shock wave is shifted outwards and at some point the steady solution ceases to 
exist. As mentioned already, however, we do not need to search that point, since the shock revival
occurs earlier owing to the hydrodynamical instabilities. We do need to identify this point, to
which we refer the critical point in this paper, in the second step.

The hydrodynamical simulations are performed both in 1D and in 2D in the second step. As mentioned 
just now, these computations are used to judge whether the spherically symmetric, steady accretion flows,
which are obtained in the first step, induce shock revival by the hydrodynamical instabilities. 
The nature of the instabilities are different between 1D and 2D: in 1D radial oscillations become
over-stabilized at some luminosity, which is lower than the one, at which the steady flow ceases to
exist~\citep{2006ApJ...641.1018O,2012ApJ...749..142F}; in 2D, on the other hand, the non-radial instability called SASI 
occurs even earlier on~\citep{2007ApJ...656.1019Y}. Hence in reality the latter will be more important. We think
that 1D models are still useful to understand the physical processes that occur after the shock relaunch 
as well as to elucidate the differences caused by the dimensionality of hydrodynamics.

\begin{figure}[ht]
\epsscale{1.0}
 \plotone{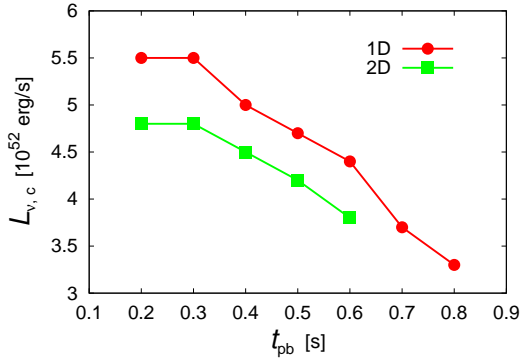}
\caption{The critical neutrino luminosities in 1D and 2D as a function of the post-bounce time.}
\label{fig:fig6}
\epsscale{1.0}
\end{figure}

We solve numerically Eqs.~(\ref{eq:cont}) - (\ref{eq:yi_flow}) with all the time derivatives turned on.
Both the input physics and radial grid are identical to those employed in the first step, in which the 
steady accretion flows are calculated. In 2D simulations we deploy $60$ grid points in the $\theta-$direction
to cover $180^{\circ}$ and add random $1$\% perturbations to the radial velocity to induce SASI. 
In these simulations we fix both the outer and inner boundary conditions 
and follow the 
evolution for $200{\rm ms}$. If the shock wave reaches the outer boundary located at $r=500 {\rm km}$ within 
this period, we judge that the shock is successfully revived. The reason why we fix the boundary conditions
is that if the shock revival occurs at a certain time, the instabilities should have reached the nonlinear
stage by that time but the growth of the instabilities takes some time. For each mass accretion rate we determine  
the minimum luminosity for the successful shock revival within a few percent and refer to it as the critical 
luminosity.
 In 1D the shock relaunch is preceded by the growth of 
radial oscillations (not shown in the figure) whereas the non-radial modes with $\ell =1, 2$ are followed by the shock revival in 2D.
Here $\ell$ stands for the index of spherical harmonics in the expansion of unstable modes. In Fig.~\ref{fig:fig6} 
the critical luminosities are presented both for 1D and 2D as a function of mass accretion rate. It is evident 
that the critical luminosity is decreasing function of mass accretion rate and it is reduced in 2D, both of which are
well known~\citep{2006ApJ...641.1018O,2008ApJ...688.1159M}.

\subsection{Step 3: computations of post-relaunch evolutions\label{sec:step3}}
In this section we give some details of the 1D and 2D hydrodynamical simulations of
post-shock-revival evolutions. We continue the computations of step 2 for the models
with the critical luminosities. We first map the results to a larger mesh that covers 
the region extending from the neutrino sphere to the radius of $r=2 \times 10^5{\rm km}$.
In all 1D models we computed the post-revival evolutions for $\sim 2 {\rm s}$, which
is found to be long enough to estimate the explosion energy. In fact, we follow the 
evolutions for two 1D models until the shock reaches the stellar surface, which is located 
at
 $r=5\times 10^8{\rm km}$.
 In those simulations we expand the mesh twice as the shock 
propagates outward. 
The inner boundary is also shifted to larger radii, to $r=10^{3}$km 
for the first re-griding and to $10^{4}$km for the second expansion, 
so that we could avoid too severe CFL conditions 
on the time step. We confirmed that these shift of 
the inner boundary do not violate 
the energy conservation in ejecta by more than 0.1\%. 
We also performed long simulations in the similar way 
(see \S\ref{sec:2d}) for three 2D models in order to determine 
the asymptotic ejecta mass accurately. 
In all simulations we employ non-uniform 
650 radial grid points in 1D and 500 points in 2D. 
In 2D simulations, 60 grid points are distributed uniformly in 
the $\theta-$direction to cover the entire meridian section.

The outer boundary condition poses no problem this time, since it is located at a very large
radius. We just impose the free in-flow/out-flow condition there. The inner boundary conditions
are a bit more difficult. We assume the time evolution of neutrino luminosity is given by
\begin{equation}
L_\nu(t_{exp}) = L_{\nu,c} \cdot \exp(- t_{exp} /800{\rm ms}),
\label{eq:levo}
\end{equation}
where $t_{exp}$ is the time elapsed from the shock relaunch and $L_{\nu,c}$ is the critical
luminosity obtained in Step~2.
 We fix the density, pressure and velocity at the ghost mesh point at
the inner boundary when matter is flowing inward. When matter begins to flow outward, i.e.
the transition to the neutrino wind phase occurs, those quantities are extrapolated from 
the innermost active mesh point to the ghost mesh point except when the entropy per baryon 
tends to be too high, in which case we put the upper bound of $s=100k_B$ to the entropy 
per baryon and the density is adjusted. These prescriptions are applied to each angular grid
point at the inner boundary for 2D simulations.

We investigate seven 1D models, for which the stalled shock is relaunched at $t_{pb} = 200$, 
$300$, $400$, $500$, $600$, $700$ and $800{\rm ms}$.
Note that $t_{pb}$ has a one to one correspondence with the mass accretion rate, which is shown in Fig.~\ref{fig:fig2}.
 Five 2D simulations are performed, in which the 
shock revival is assumed to occur at $t_{pb}=200$, $300$, $400$, $500$ and $600{\rm ms}$.
See Fig.~\ref{fig:fig6} for the critical luminosities in these models. The input physics, such
as nuclear and weak interactions, are the same as those employed in the second step of Step~2.
The results of all the computations in this step will be presented in the next section first for 
the 1D models and then for the 2D cases.

\section{Results}
\label{sec:results}
\subsection{Spherically symmetric 1D models}
\label{sec:1d}
\subsubsection{The evolution of the fiducial model}
We first describe in detail the evolution of the 1D model, in which the 
shock relaunch is assumed to occur at $t_{pb}=400{\rm ms}$. This corresponds to
the time, at which the mass accretion rate is 
$0.53{\rm M}_{\odot}$/s. 
The critical
luminosity in 1D is $5\times10^{52}{\rm erg/s}$.

\begin{figure*}
\vspace{15mm}
\epsscale{1.2}
\plotone{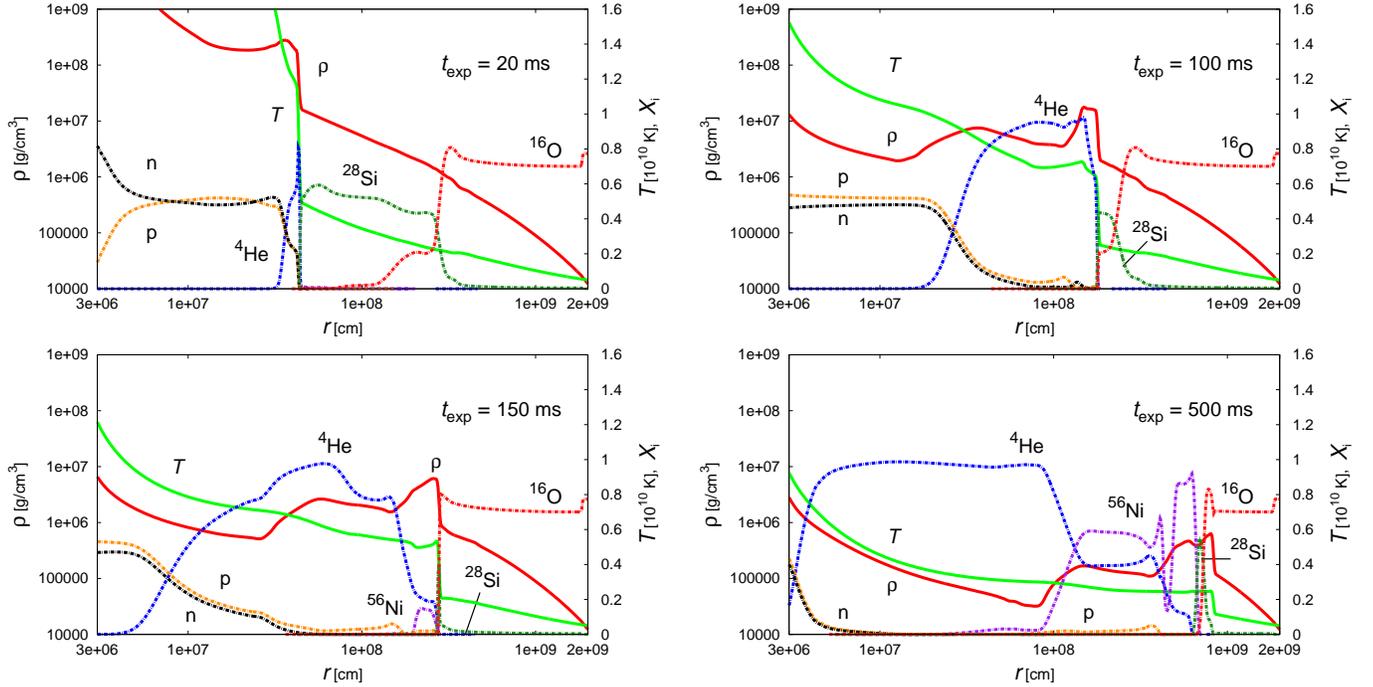}
\caption{The density (red solid line), temperature (green solid line) 
and mass fractions of representative nuclei (solid dotted lines) 
for the 1D fiducial model as a function of radius at four different post-relaunch times, $t_{exp}=$20, 100, 150 and 500ms, respectively. 
The last snapshots show $^{56}{\rm Ni}$ production (purple) 
and $\alpha$-rich freeze-out (blue). One can also see in the last panel 
the $^{28}{\rm Si}$ production (dark green) by the O-burning (magenda). 
\label{fig:fig7}}
\epsscale{1.2}
\end{figure*}


In Fig.~\ref{fig:fig7} we show the density, temperature and mass fractions of
representative nuclei as a function of radius for four different times. 
In the upper left panel the profile at
$t_{exp}=20{\rm ms}$
 after the shock relaunch is displayed.
 The shock is still located around $r=500{\rm km}$. The post-shock temperature is 
$T \gtrsim 1{\rm MeV}$, so high that the nuclei, mainly $^{28}{\rm Si}$, flowing into 
the shock are decomposed to $\alpha$ particles, which are further disintegrated into
nucleons immediately. The post shock composition is perfectly described by NSE. At
$t_{exp}=100{\rm ms}$
 the shock reaches $r \sim 2,000{\rm km}$ but is still inside the $\rm{Si}$ 
layer as seen in the upper right panel. The post-shock matter is mainly composed of 
$\alpha$ particles, which are not disintegrated any more owing to the lower temperature, 
$T \sim 7\times10^9{\rm K}$. Another $50{\rm ms}$ later, the shock enters the Oxygen layer
(see the lower left panel). Now $^{56}{\rm Ni}$ emerges just behind the shock wave. 
This is mainly due to the recombination of $\alpha $ particles, which will be evident shortly.
The post-shock temperature is $T\sim 5\times 10^9{\rm K}$ and matter is beginning to 
be out of NSE. In the lower right panel we present the profile at
$t_{exp}=500 {\rm ms}$.
At this time, the temperature is $T\sim 2\times 10^9{\rm K}$ and matter is completely
out of NSE and the nuclear reactions yield mainly $^{28}{\rm Si}$. Much behind the shock
wave some $\alpha$ particles are recombined to $^{56}{\rm Ni}$. Slightly later all
nuclear reactions are terminated behind the shock wave, since the temperature does not 
rise to high enough values by shock heating.

\begin{figure}[ht]
\epsscale{1.0}
 \plotone
{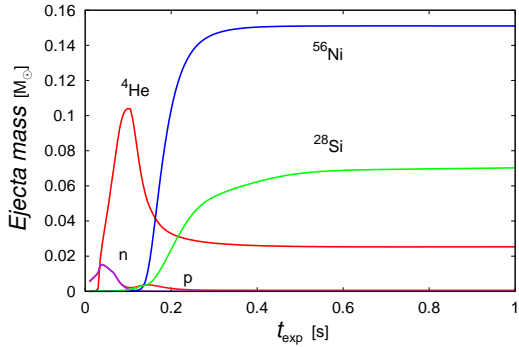}
\caption{The ejecta masses of proton (dark red line)
, neutron (purple line), $\alpha$ (red line)
, $^{28}{\rm Si}$ (green line) and $^{56}{\rm Ni}$ (blue line) 
integrated over the post-shock region are displayed as a function of 
the elapsed time, $t_{exp}$, for the 1D fiducial model. 
No fall back occurs and the masses of $^{56}{\rm Ni}$ and 
$^{28}{\rm Si}$ are determined as early as $t_{exp} \sim 500{\rm ms}$.}
\label{fig:fig8}
\epsscale{1.0}
\end{figure}

In Fig.~\ref{fig:fig8} we show the masses of proton, neutron, $\alpha$, $^{28}{\rm Si}$
and $^{56}{\rm Ni}$ integrated over the region inside the shock wave as a function of 
$t_{exp}$
 for the 1D fiducial model. In accord with the description in the previous
paragraph, $\alpha $ particles are the main yield of nuclear reactions up to
$t_{exp}\sim 100{\rm ms}$.
The depletion of neutrons after
$t_{exp} \sim 50{\rm ms}$
 implies that the nucleons are recombined 
to $\alpha $ particles during this period.
From $t_{exp} \sim 100{\rm ms}$ to $t_{exp} \sim 150 {\rm ms}$, 
on the other hand, $\alpha $ particles are diminished while $^{56}{\rm Ni}$ is increase, which 
means that the former is recombined to the latter. After
 $t_{exp} \sim 150{\rm ms}$ $\alpha $
 particles
cease to recombine any more and are frozen, and $^{56}{\rm Ni}$ and later $^{28}{\rm Si}$ 
are produced by nuclear burnings. These results are obtained with the nuclear network with 
28 nuclei (see \S\ref{sec:method}). In order to confirm that it is large enough, 
we conduct a larger network with 463 nuclei
as a post-processing calculation, employing the time evolutions of density, 
temperature and electron fraction obtained by the simulation with the original network.
The nickel and silicon masses are $0.140{\rm M}_\odot$ and $0.068{\rm M}_\odot$ for the larger network, whereas they are $0.151{\rm M}_\odot$ and $0.071{\rm M}_\odot$ for the standard case. 
Further more, the difference in the total mass of heavy nuclei with
$\rm A \ge 48$ is only $2.0 \times 1.0^{-3} {\rm M}_\odot$ 
and the additional energy release from these difference 
is estimated to be less than $1.0 \times 10^{49}$erg. 
These results imply that
the original network is appropriate.

\begin{figure}[ht]
\epsscale{1.0}
 \plotone
{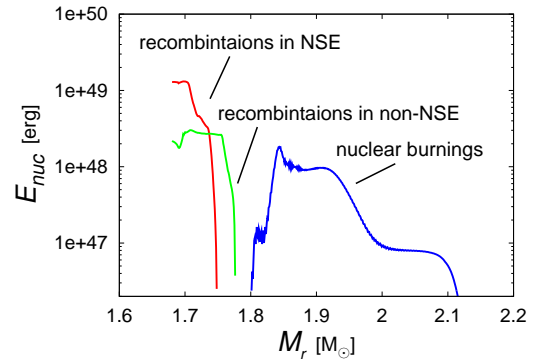}
\caption{Total energy release by nuclear reactions as 
a function of mass coordinates. 
The size of mass bin is 
$1.0 \times 10^{-3} {\rm M}_\odot$. 
The energy production rate is given by integrating over the entire evolution. 
mass bins being taken as 
Three contributions, i.e., recombinations in NSE (red line), those 
in non-NSE (green line) and nuclear burnings (blue line), are distinguished. 
}
\label{fig:fig9}
\epsscale{1.0}
\end{figure}

The energy release by these nuclear reactions is presented in Fig.~\ref{fig:fig9}. We time-integrate the
energy production rate over the entire evolution as a function of mass coordinates. 
In so doing, we distinguish the contributions of the recombinations from those of the nuclear burnings. Moreover,
the former is divided into two pieces, one of which comes from the recombinations in NSE and the other
from those in non-NSE. In the interior ($\lesssim 1.75{\rm M}_\odot$) the recombinations start in NSE
and end in non-NSE and there are hence two contributions. In the slightly outer layer up to $\sim 1.8{\rm M}_\odot$,
on the other hand, the recombinations occur in non-NSE conditions. Further outside ($\gtrsim 1.8{\rm M}_\odot$) 
the nuclear burnings take their places. 
In fact, the densities and temperatures that the matter 
in this region expansions are inside the O-burning regime 
(see Fig.1 in \citet{Hix&Thielemann(1999)}). 
It is evident that the largest energy release comes from the 
recombinations that occur in NSE and the contributions of the nuclear burnings are rather minor even
after the integrated over the mass coordinate for this particular model. This is a generic trend as will be
shown later. There is a gap between $\sim 1.75{\rm M}_\odot$ and $\sim 1.8{\rm M}_\odot$. This is a region,
in which energy is not released but absorbed. The main reaction in this region is the burning of $^{28}{\rm Si}$
to $^{56}{\rm Ni}$. Some fractions of $^{28}{\rm Si}$ are disintegrated to $\alpha$ particles, however. Although
the latter is minor, the mass difference between $\alpha$ and $^{28}{\rm Si}$ is much greater than that between
$^{28}{\rm Si}$ and $^{56}{\rm Ni}$. As a results the energy suck by the decomposition overwhelms the energy release 
by the burning.

\subsubsection{The evolution of diagnostic explosion energy \label{sec:diag}}

Understanding the evolutions of density, temperature, chemical composition as well as the energy generations by
nuclear reactions, we now turn our attention to the explosion energy. Following the conventional practice, we
define the diagnostic explosion energy of provisional ejecta. At each grid point the total energy 
density, $e_{tot}$, is given by 
\begin{equation}
e_{tot} = e_{kin} + e_{int} + e_{grav},
\label{eq:etot}
\end{equation} 
where $e_{kin}= 1/2 \rho v^2$ is the kinetic energy density, $e_{int}$ denotes the internal energy density and
$e_{grav}= \rho (\Phi + \Phi_c)$ stands for the gravitational potential energy density. We judge that the mass element
at a certain grid point will be ejected if the total energy density is positive, $e_{tot}>0$, and if the radial 
velocity is positive ($v_{r}>0$) at a given time. Then the diagnostic explosion energy is defined as a function of
time to be the sum of the total energy density times volume over the ejecta, which is just determined. 

The diagnostic explosion energy changes in time indeed. In the early phase of shock revival, the neutrino heating
is the main source of the diagnostic explosion energy. As the shock propagates outward, the neutrino heating becomes
inefficient, since the matter to be heated is also shifted to larger radii, where the neutrino flux is lower, and
the luminosity itself becomes smaller as the time passes. It is also important that nucleons, which are mainly
responsible for the heating, are depleted as they recombine to $\alpha$ particles and heavier nuclei as the temperature 
decreases. After the neutrino heating subsides, the nuclear reactions are the main energy source. As described in the
previous section, the recombination of nucleons occurs at first and the nuclear burnings take their place later.
After all nuclear reactions are terminated owing to low temperatures, the diagnostic explosion energy decreases
slowly since matter, which is gravitationally bound and hence has negative specific energy, is swallowed by the 
shock wave. As the shock wave proceeds outwards, this contribution becomes smaller and the diagnostic explosion
energy approaches its asymptotic value, the actual explosion energy.

\begin{figure}[ht]
\epsscale{1.0}
 \plotone
{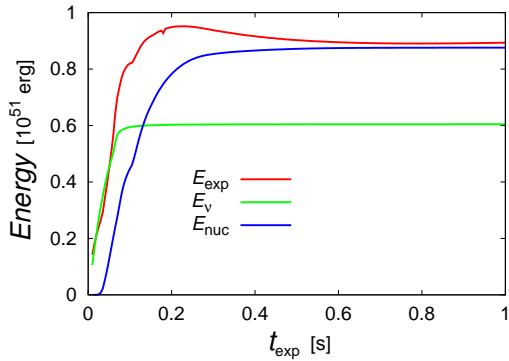}
\caption{The time evolution of diagnostic explosion energy for the 1D fiducial model. The individual contributions
from the neutrino heating (green line) as well as nuclear reactions (blue line) are also shown.}
\label{fig:fig10}
\epsscale{1.0}
\end{figure}

Figure~\ref{fig:fig10} shows the time evolution of the diagnostic explosion energy for the 1D fiducial model, in
which the shock is relaunched at $t_{pb}=400{\rm ms}$. The horizontal axis in the figure is
the time elapsed from 
the shock relaunch, $t_{exp}$. The diagnostic explosion energy increases for the first $\sim 200{\rm ms}$. Then it decreases 
gradually and becomes almost constant at
$t_{exp} \sim 1$s.
 Also displayed in the figure are the individual 
contributions to the diagnostic explosion energy from the neutrino heating and nuclear reactions. As described 
in the previous paragraph, the neutrino heating is dominant over the nuclear reactions initially up to
$t_{exp} \sim 120{\rm ms}$.
Then the nuclear reactions become more important and raise the diagnostic explosion energy to $\sim 10^{51}{\rm erg}$
at by the time
$t_{exp} \sim 200{\rm ms}$
 in this particular case. As indicated by the colored shades in the figure, 
the nuclear reactions are mainly the recombinations until
$t_{exp} \sim 150{\rm ms}$.
 The nuclear burnings follow 
until
$t_{exp} \sim 300{\rm ms}$.
 The asymptotic value of the diagnostic explosion energy is approached from above
owing to the engulfing of matter with negative energy by the outgoing shock wave.

\begin{figure}[ht]
\epsscale{1.0}
 \plotone
{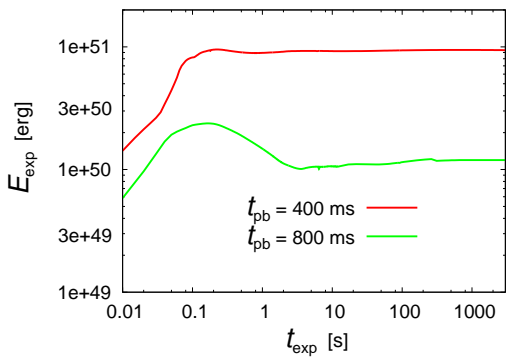}
\caption{The long-term evolutions of diagnostic explosion energies for the 1D fiducial model as well as
for the model of the latest shock relaunch. }
\label{fig:fig11}
\epsscale{1.0}
\end{figure}

In order to confirm that the final explosion energy has been already reached in the above computation, we continue 
to evolve this model until the shock wave reaches the stellar surface. We shift both the outer and inner boundaries 
as mentioned in \S\ref{sec:step3} to avoid too severe CFL conditions at the innermost mesh point,. The result is presented in 
Fig.~\ref{fig:fig11}. It is clear that the diagnostic explosion energy is essentially constant for
$t_{exp} \gtrsim 1{\rm s}$.
Also shown in the figure is the result for another model, in which the shock relaunch is delayed until $t_{pb}=800{\rm ms}$.
The accretion late is $\sim 0.23{\rm M}_{\odot}/{\rm s}$ and the critical luminosity is $\sim 3.3\times 10^{52}{\rm erg/s}$ 
for this 1D model. As is obvious from the figure, the asymptotic explosion energy is considerably smaller, 
$\sim 1.1\times 10^{50}{\rm erg}$, and we have to wait for $\sim 2{\rm s}$ before the diagnostic explosion 
energy is settled to the asymptotic value. This is a generic trend: as the shock relaunch is delayed, it takes more 
time to reach the final explosion energy.

\subsubsection{Systematics \label{sec:sys}}

\begin{figure}[ht]
\epsscale{1.0}
 \plotone
{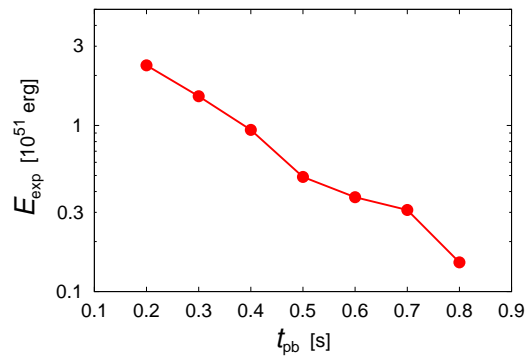}
\caption{The asymptotic values of diagnostic explosion energy for all 1D models.}
\label{fig:fig12}
\epsscale{1.0}
\end{figure}

In this section, we look into the results of other models and see how generic what we have found so far for the fiducial model
is. In Fig.~\ref{fig:fig12} we show the asymptotic values of diagnostic explosion energy for different models as a 
function of the shock-relaunch time. We can clearly see that the explosion energy is a monotonically decreasing function
of the shock-revival time. This is main due to the fact that the mass of accreting matter gets smaller as the time passes
and is nothing unexpected~(see e.g. \citep{2006A&A...457..963S}). It is stressed, however, this is the first clear demonstration of the fact 
with the nuclear reactions and EOS being taken into account consistently in sufficiently long computations, in which the diagnostic
explosion energy is confirmed to reach the asymptotic value.

Also shown in the figure are the individual contributions to the diagnostic explosion energy from the nuclear reactions and 
neutrino heating. Both of them also decrease as the shock revival is delayed. It is found, however, that the contribution of
the nuclear reactions diminishes more rapidly. This is simply due to the fact that the temperature rise by the shock passage
is smaller in weaker explosions. Note that the explosion energy are smaller than the sum of the two contributions, since
the accretion of gravitationally bound matter gives negative contributions to the explosion energy as mentioned already. 
It should be also emphasized that the recombination energy is eventually originated from the neutrino heating because the 
recombinations are necessarily preceded by the endothermic dissociations of heavy nuclei that exist prior to collapse and 
those consumed energies are replenished by neutrinos. The neutrino heating also plays a vital role to push the post-bounce
configuration to the critical point and further heat up matter until they become gravitationally unbound in the earliest phase of 
shock revival.

\begin{figure}[ht]
\epsscale{1.0}
 \plotone
{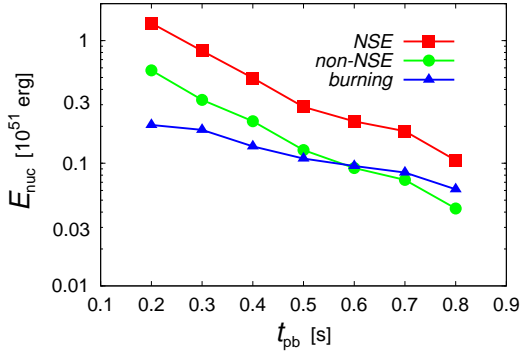}
\caption{The individual contributions from the nuclear recombinations in NSE and those in non-NSE as well as nuclear burnings to the explosion energy.}
\label{fig:fig13}
\epsscale{1.0}
\end{figure}

In Fig.~\ref{fig:fig13} we further divide the contribution of nuclear reactions into those from the recombinations in and out of
NSE as well as from the nuclear burnings. Roughly speaking, the re-assemble of nucleons to $\alpha $ particles occurs in the
recombination in NSE and the further recombinations to heavier nuclei proceed in the environment out of NSE. As can be seen, 
all the contributions again decline as the shock relaunch is delayed. Regardless the recombination that occurs in NSE is 
the greatest contributor. The recombinations, both in and out of NSE, decline more rapidly than the nuclear burning and 
the latter contributes more than the recombination out of NSE for the model, in which the shock revival occurs at the 
latest time ($t_{pb}=800 {\rm ms}$) and the weakest explosion is obtained. The reason why the nuclear burning declines
more slowly is that the temperatures obtained by shock heating is roughly proportional to the quarter power of the explosion
energy.

\begin{figure}[ht]
\epsscale{1.0}
 \plotone
{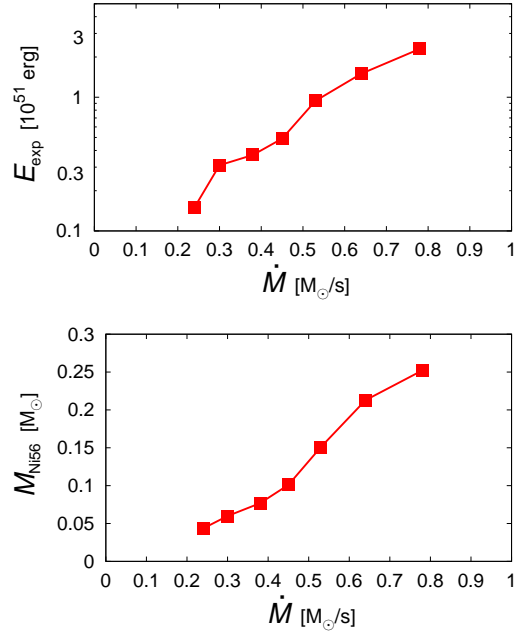}
\caption{The explosion energies and $^{56}{\rm Ni}$ masses for all 1D models. }
\label{fig:fig14}
\epsscale{1.0}
\end{figure}

Next we turn our attention to the synthesis of $^{56}{\rm Ni}$, one of the most important observables in the supernova explosion.
The synthesized mass of $^{56}{\rm Ni}$ is correlated with the explosion energy~\citep{2003ApJ...582..905H}: the greater the explosion energy
is, the more $^{56}{\rm Ni}$ is produced. This is demonstrated in Fig.~\ref{fig:fig14}. In the upper panel we again present the 
asymptotic values of the diagnostic explosion energy, which we simply refer to as the explosion energy here, as a function of 
the mass accretion rate at shock relaunch. The corresponding shock-relaunch times are given in the figure. In the lower panel,
the mass of $^{56}{\rm Ni}$ in the ejecta is displayed also as a function of the mass accretion rate at shock relaunch. The ejecta
was defined earlier to be the collection of the mass elements that have positive total energy density and radial velocity 
(see \S\ref{sec:diag}). The positive correlation between the explosion energy and the mass of $^{56}{\rm Ni}$ in the ejecta 
are evident.

In fact, $^{56}{\rm Ni}$ may be produced too much in these 1D models. 
The canonical 
explosion energy ($\sim 10^{51}{\rm erg}$) is attained only by the models that relaunch the stalled shock wave relatively 
early ($t_{pb} \lesssim 400{\rm ms}$).
On the other hand, the masses in the ejecta of $^{56}{\rm Ni}$ synthesized for all these models are $\gtrsim 0.15{\rm M}_{\odot}$, 
which is substantially larger than the values estimated from observations, $\lesssim 0.1{\rm M}_{\odot}$~\citep{2009MNRAS.395.1409S}. Note that 
the mass of $^{56}{\rm Ni}$ ejected by SN1987A is estimated to be $\sim 0.07{\rm M}_{\odot}$ and corresponds to the shock-relaunch
time of $t_{pb} \sim 600 {\rm ms}$ in our 1D model; this rather late shock revival gives only a weak explosion of 
$\sim 0.4 \times 10^{51}{\rm erg}$,
smaller than the most likely explosion energy ($0.9\times 10^{51}{\rm erg}$) derived observationally~\citep{2009ApJ...703.2205K}.
 It is true that
both of the observational estimates and the theoretical predictions presented here have uncertainties. In fact, the neutrino
transport as well as evolutions of proto-neutron star, which are neglected and roughly mimicked in this paper, are the main source
of uncertainties in the results shown above. We believe, however, that the general trends would be unchanged even if more sophisticated
treatments were adopted. The above argument hence may be regarded as yet another reason that we do not believe that the 1D neutrino
heating works. 

\begin{table*}
\begin{center}
\caption{Comparison with an ordinary explosive-nucleosynthesis calculation.}
\begin{tabular}{ccccccc}\hline
Shock relaunch time& \multicolumn{2}{c}{Explosion energy} & 
\multicolumn{2}{c}{Proto-neutron star mass} &\multicolumn{2}{c}{$^{56}{\rm Ni} $ mass}\\
\multicolumn{1}{c}{$[{\rm ms}]$} & \multicolumn{2}{c}{$[10^{51}{\rm erg}]$} & \multicolumn{2}{c}{$[{\rm M}_{\odot}]$} & \multicolumn{2}{c}{$[{\rm M}_{\odot}]$}\\ \hline
& 1D model & thermal bomb & 1D & thermal bomb & 1D & thermal bomb\\ \cline{2-7} 
300 & 1.50 & 1.50 & 1.60 & 1.59 & 0.21 & 0.15 \\
400 & 0.95 &0 .94 & 1.67 & 1.64 & 0.15 & 0.086 \\
500 & 0.50 & 0.60 & 1.70 & 1.73 & 0.10 & 0.043 \\ \hline
\end{tabular}
\end{center}
\label{tab:tab1}
\end{table*}

\begin{figure}[ht]
 \plotone{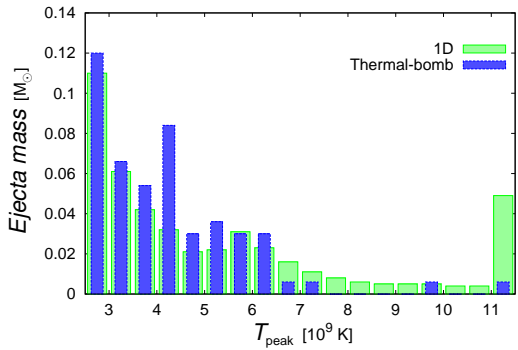}
\caption{Comparison of the distributions of peak temperature between the 1D fiducial model and corresponding thermal-bomb model. The total mass 
of the matter that has the peak temperature higher than $T_{9}=5$ is 
$1.89 \times 10^{-1} {\rm M}_{\odot}$ for the 1D model whereas it is
$1.20 \times 10^{-1} {\rm M}_{\odot}$ for the themal-bomb model.}
\label{fig:fig15}
\end{figure}

To better understand the origin of the overproduction of $^{56}{\rm Ni}$ we have done an ordinary calculation of explosive nucleosynthesis
as a post-process for the densities, temperatures and electron fractions obtained for the models presented above. In so doing,
the so-called thermal bomb method, in which thermal energy is deposited initially in the innermost region, 
is employed~\citep{hashimoto95}. 
The explosion energy and mass cut are chosen so that they agree with those of the original models. Interestingly the calculation of
explosive nucleosynthesis consistently produces smaller amounts of $^{56}{\rm Ni}$, which is given in Table~\ref{tab:tab1}. In fact, 
the fiducial model reproduces the observational estimate for SN1987A much better, which is just a coincidence though. Figure~\ref{fig:fig15}
compares the distributions of peak temperature between the 1D fiducial model and corresponding thermal-bomb model. It is clear that
the fiducial model has a larger amount of mass elements that achieve high enough temperatures to produce $^{56}{\rm Ni}$. It seems
to take the neutrino heating mechanism a greater thermal energy to unbound the accreting envelope. This result may also be a caution in
employing the thermal bomb method in the explosive nucleosynthesis calculations.

In fact, 
\citet{young07} discussed uncertainty in nucleosynthetic yields with 1D spherically symmetric explosion models
and showed that \nuc{Ni}{56} evaluated with the piston-driven model ($E_{\rm exp}$ = 1.2 $\times 10^{51}$erg) 
is larger by a factor of 2-3 than that with the thermal-bomb model ($E_{\rm exp}$ = 1.5 $\times 10^{51}$erg) with the same remnant mass.
Our 1D neutrino-driven explosion models may be closer to the piston-driven explosion model than to the thermal-bomb explosion model.

\subsection{Axisymmetric 2D models}
\label{sec:2d}
In the following we focus on the effect of dimension. In \S\ref{sec:step2} we already 
showed that the critical neutrino luminosity is lowered in the axisymmetric 2D models
than in the spherically symmetric 1D model for the same accretion rate at the shock 
relaunch. Performing 2D simulations further in Step 3, we are concerned with how and how much
the results we obtained for 1D models so far are modified in 2D cases.  

The input physics for 2D simulations is essentially the same as for the 1D model except for 
the perturbations added to the radial velocities with random magnitudes up to $1$\% at the beginning of
Step 2. We have investigated the models, in which the stalled shock wave is relaunched at the post-bounce
times of
$t_{pb} = 200, 300, 400, 500$ and $600$ms.
For the models with $t_{pb}=400, 500$ and 600ms, 
we further extend the domain twice later up to $r=2 \times 10^{7}$km as already mentioned. This is necessary to determine 
the mass of the matter that eventually falls back. 
It turns out that the model 
with $t_{pb}=$600ms fails to explode, with all the shocked matter starting to 
falling back before the shock wave reaches the He-layer. 
We also tested the numerical convergence in the model with 
$t_{pb}=$400ms by increasing the numbers of radial or angular 
grid points. We found that the higher ridial resolution 
(700 mesh points instead of 500) 
does not make much difference. Incidentally, this is also the case in 1D. 
On the other hand, the finer angular mesh 
(90 grid points instead of 60) yields 
the explosion energy and $^{56}$Ni that are
, respectively, 16\% and 10\% larger than 
those for the standard resolution.

\subsubsection{Dynamics of aspherical shock revival}
\label{sec:explosion-2d}


\begin{figure*}
\vspace{15mm}
\epsscale{0.5}
\plotone{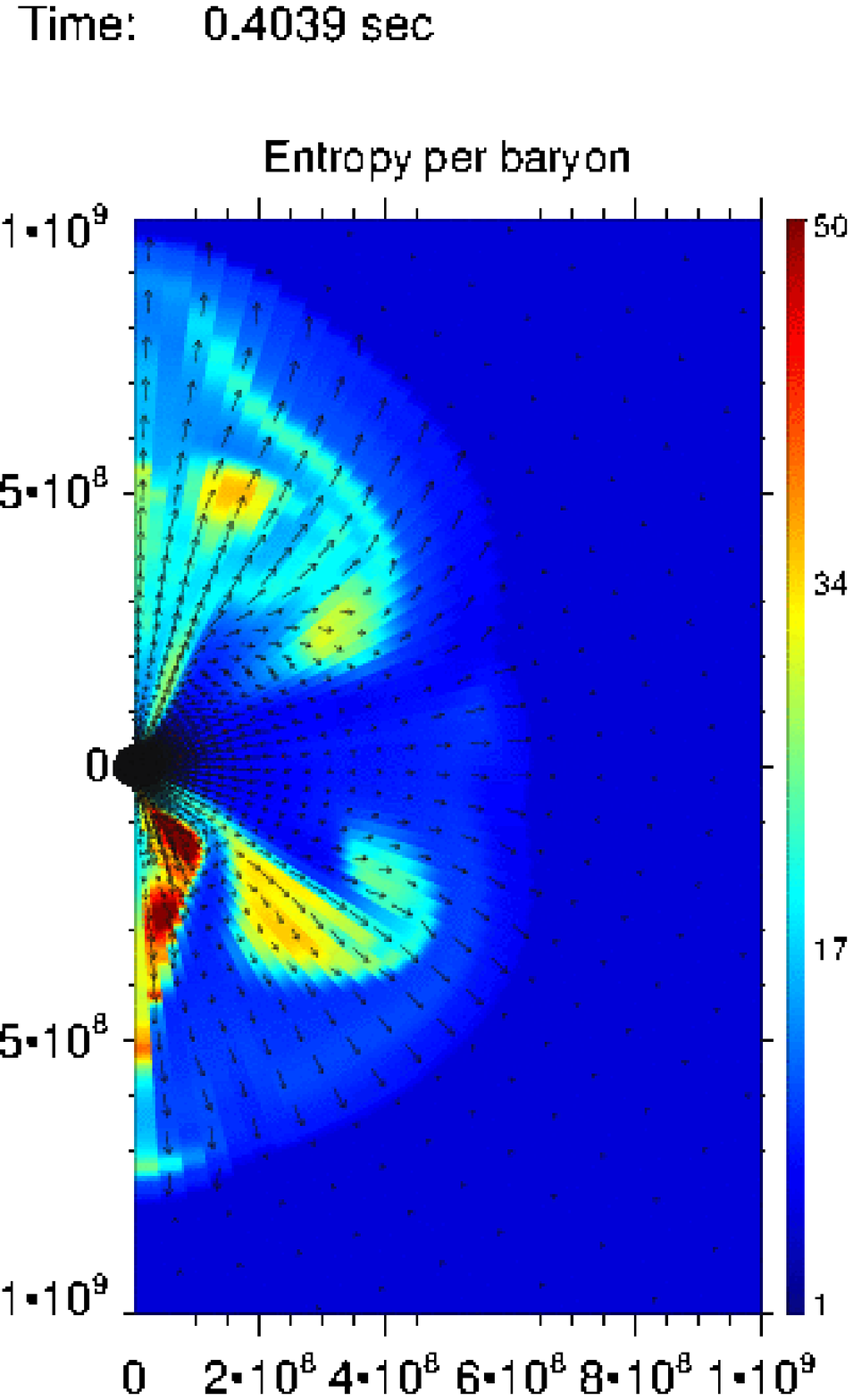}
\plotone{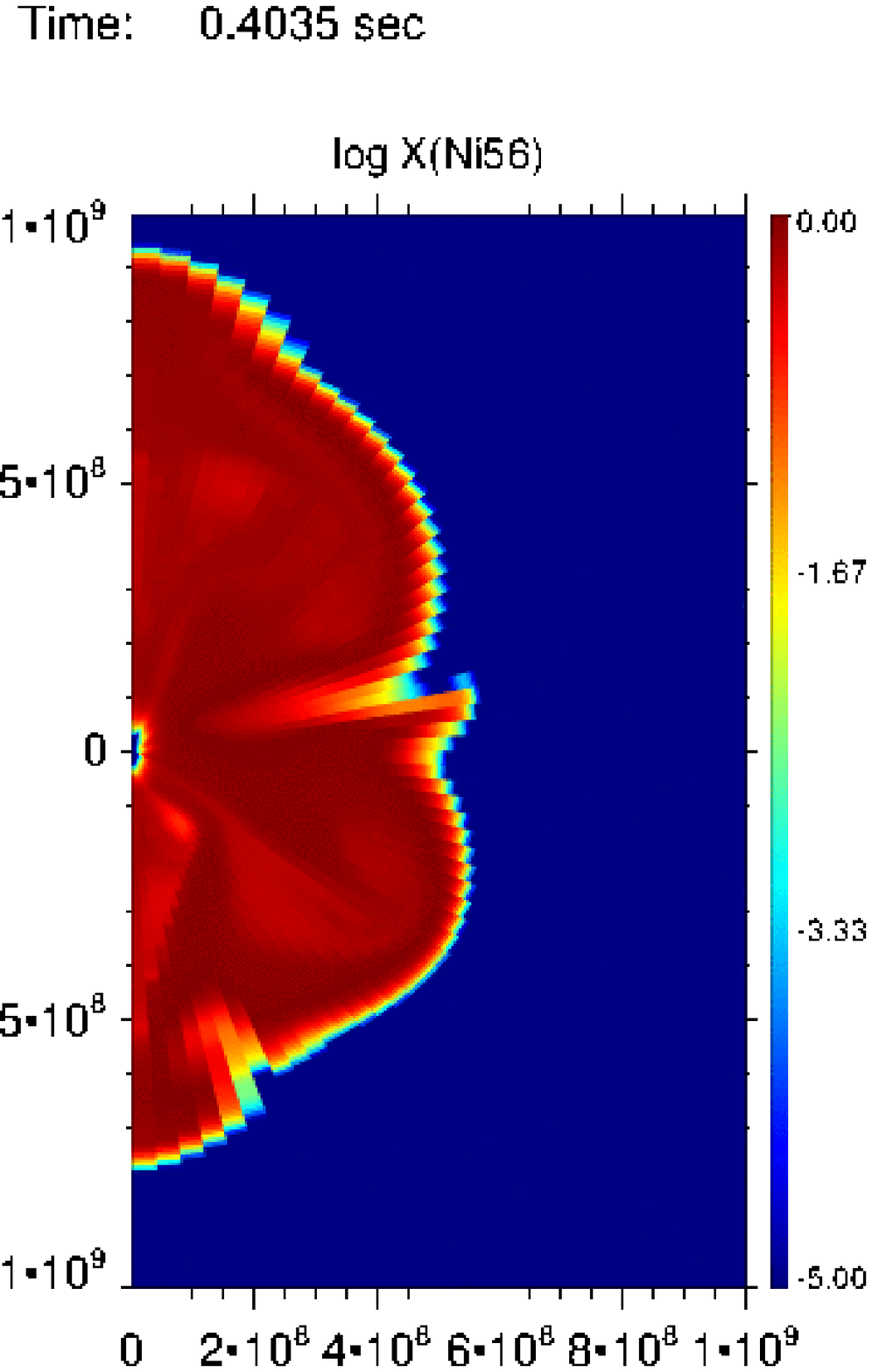}
\caption{Contours of entropy (left panel) and mass fraction of \nuc{Ni}{56} (right panel) at
$t_{exp} = 400$ms
for the 2D model, in which the stalled shock revives at
$t_{pb} = 400$ms.
\label{fig:cont-t400ms-2d}}
\epsscale{0.5}
\end{figure*}

We first look at the post-relaunch dynamics of the 
$t_{pb} = 400$ms model, 
which is a 2D counterpart to the 1D fiducial model. 
As shown in 
Fig.~\ref{fig:cont-t400ms-2d}, which displays the contours of entropy per baryon and mass fraction of 
\nuc{Ni}{56}, the shock expansion is highly aspherical, which was also demonstrated in 
\citet[e.g.]{2006A&A...453..661K,2006ApJ...641.1018O,2006A&A...457..963S}. The shock front is elongated 
in the direction of the symmetry axis. The shock propagates more rapidly in the northern hemisphere whereas
matter attains higher entropy per baryon in the opposite hemisphere. These features conform with the 
dominance of $\ell=1, 2$ modes owing to SASI. Here $\ell$ stands for the index of Legendre polynomials,
which are included in the eigenfunctions of linearly unstable modes.

\begin{figure*}
\vspace{15mm}
\epsscale{0.5}
\plotone{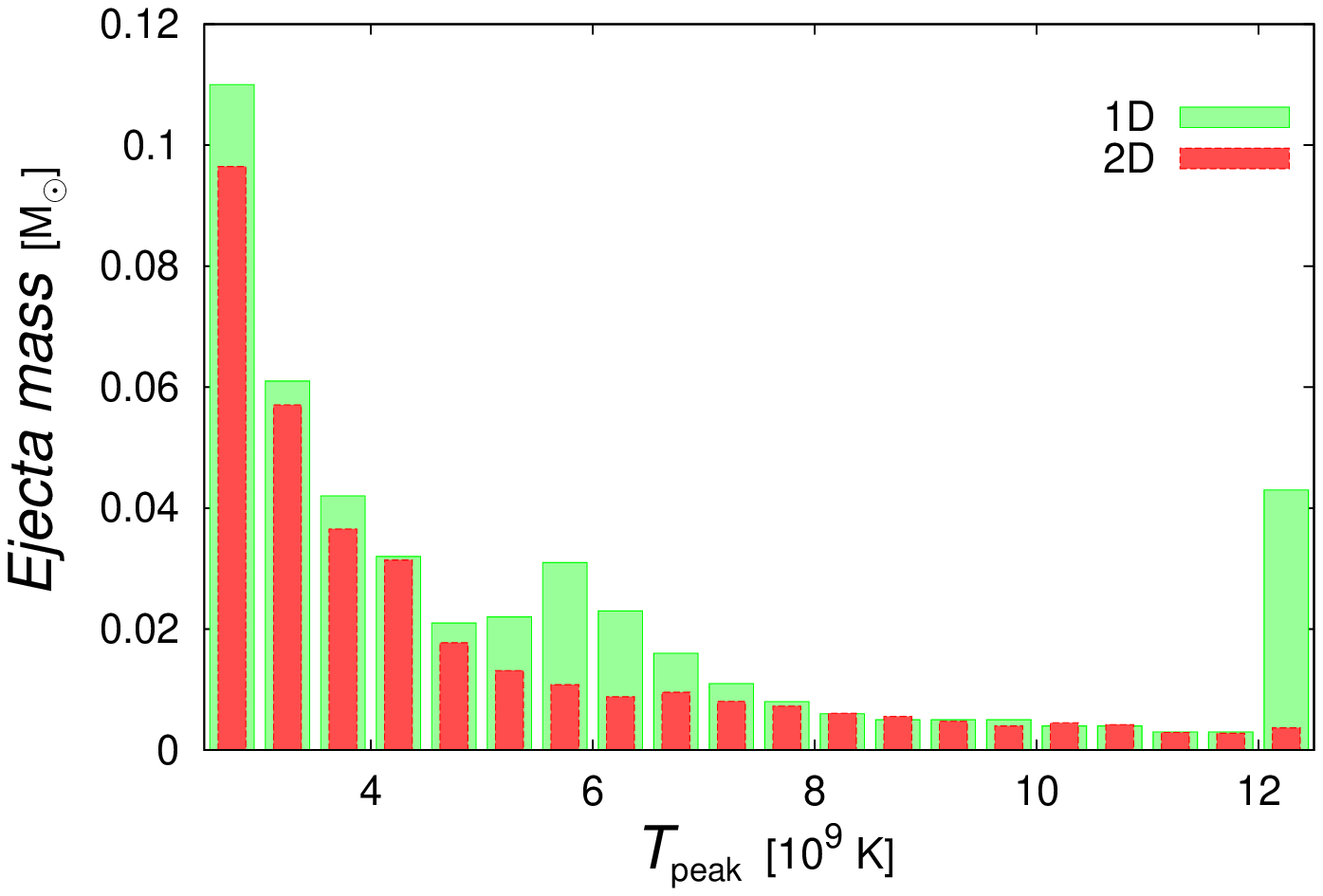}
\plotone{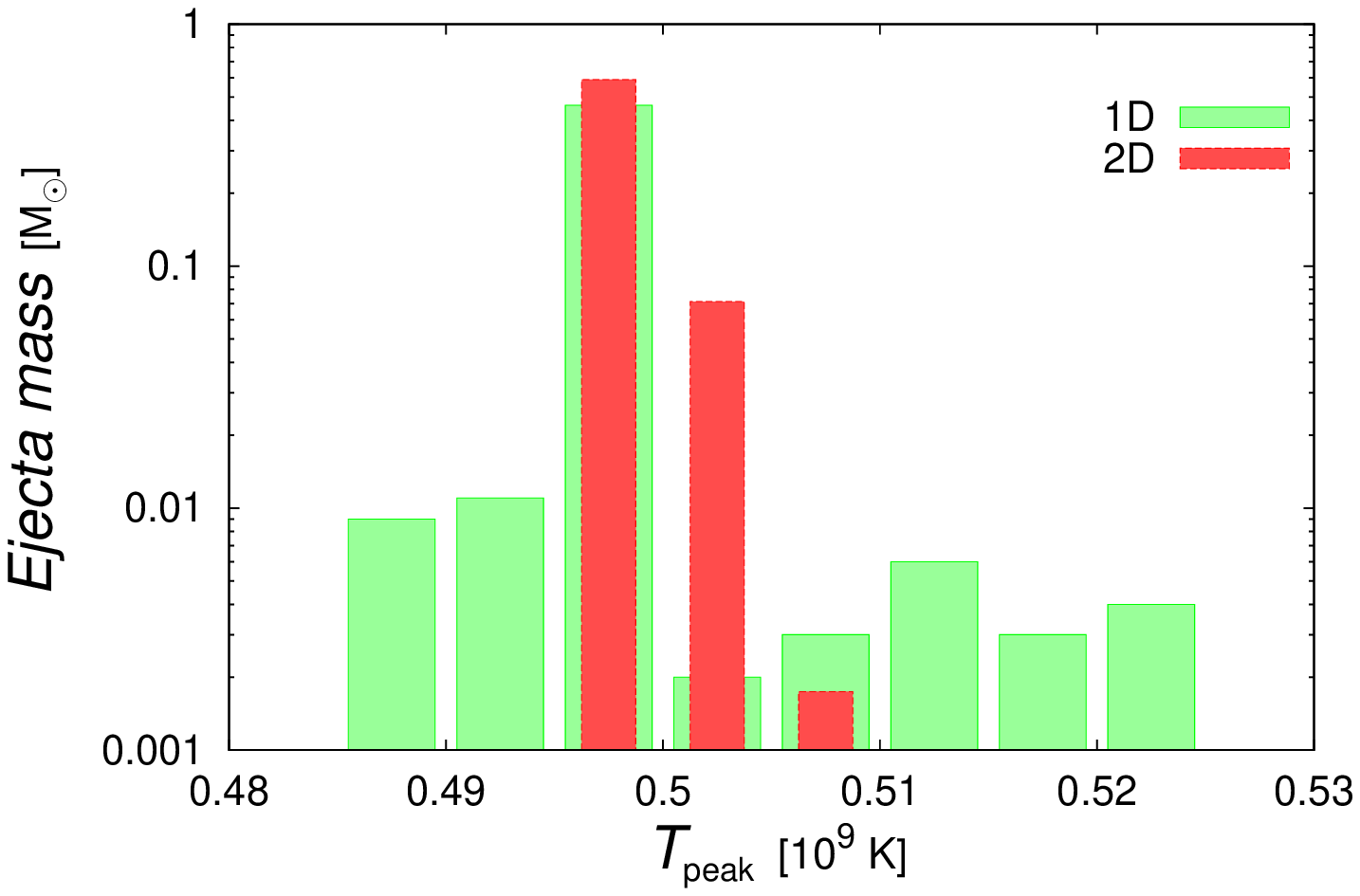}
\caption{Comparison of the distributions of peak temperatures $T_{\rm peak}$ (left panel) and 
$Y_e ({\rm NSE})$ (right panel) between the 1D fiducial model and the 2D counterpart.
The total mass of the matter that has the peak temperature higher than $T_{9}=5$ is 
$1.89 \times 10^{-1} \rm M_{\odot}$ in 1D and 
$9.62 \times 10^{-2} \rm M_{\odot}$ in 2D.
\label{fig:maxtemp_2d}}
\epsscale{0.5}
\end{figure*}



As mentioned in \S\ref{sec:step2}, the shock revival occurs at a lower luminosity in the 2D model 
($L_{c}=4.2\times10^{52}{\rm erg/s}$) than in the 1D counterpart ($L_{c}=5.0\times10^{52}{\rm erg/s}$).
Unlike the 1D case, some matter continues to accrete, forming down drafts particularly in the equatorial 
region, until much late times after the shock revival. 
As a consequence the (baryonic) mass of neutron 
star is larger in the 2D case ($\sim 2.1{\rm M}_{\odot}$) than in the 1D case ($\sim 1.65{\rm M}_{\odot}$). 
This is actually a generic trend as shown later (see Fig.~\ref{fig:pnsmass}). Another interesting feature
found in the 2D model is the distribution of maximum temperatures that each mass element attains, which
is obtained from the Lagrangian evolutions of tracer particles distributed in the ejecta. 
Figure~\ref{fig:maxtemp_2d} shows the result in a histogram. It is evident that the mass that reaches 
$T=5\times10^9{\rm K}$ is larger in 1D than in 2D. In the same figure we also show the distribution of 
electron fraction, $Y_e({\rm NSE})$, 
which is estimated when $T$ becomes the boundary value of NSE, or $7 \times 10^9 \K$.
Note that $Y_e({\rm NSE})$ is useful for the diagnosis of nuclear yields in ejecta.
In the 1D case, there exist too massive ejecta with $Y_e({\rm NSE}) \le 0.49$,
which will produce unacceptable amount of neutron-rich Ni isotopes and \nuc{Zn}{64} 
compared with the solar abundances as shown in \citet{fujimoto11}.
Their overproduction of the slightly neutron-rich ejecta with $Y_e({\rm NSE}) \le 0.49$ 
in the 1D case disappears in the 2D case owing to more efficient neutrino interactions. 
Multi-dimensional models are therefore preferable 
in the point of view of the Galactic chemical evolution of isotopes
although they may depend on the treatment of neutrino transfer.


\begin{figure*}
\vspace{15mm}
\epsscale{0.3}
\plotone{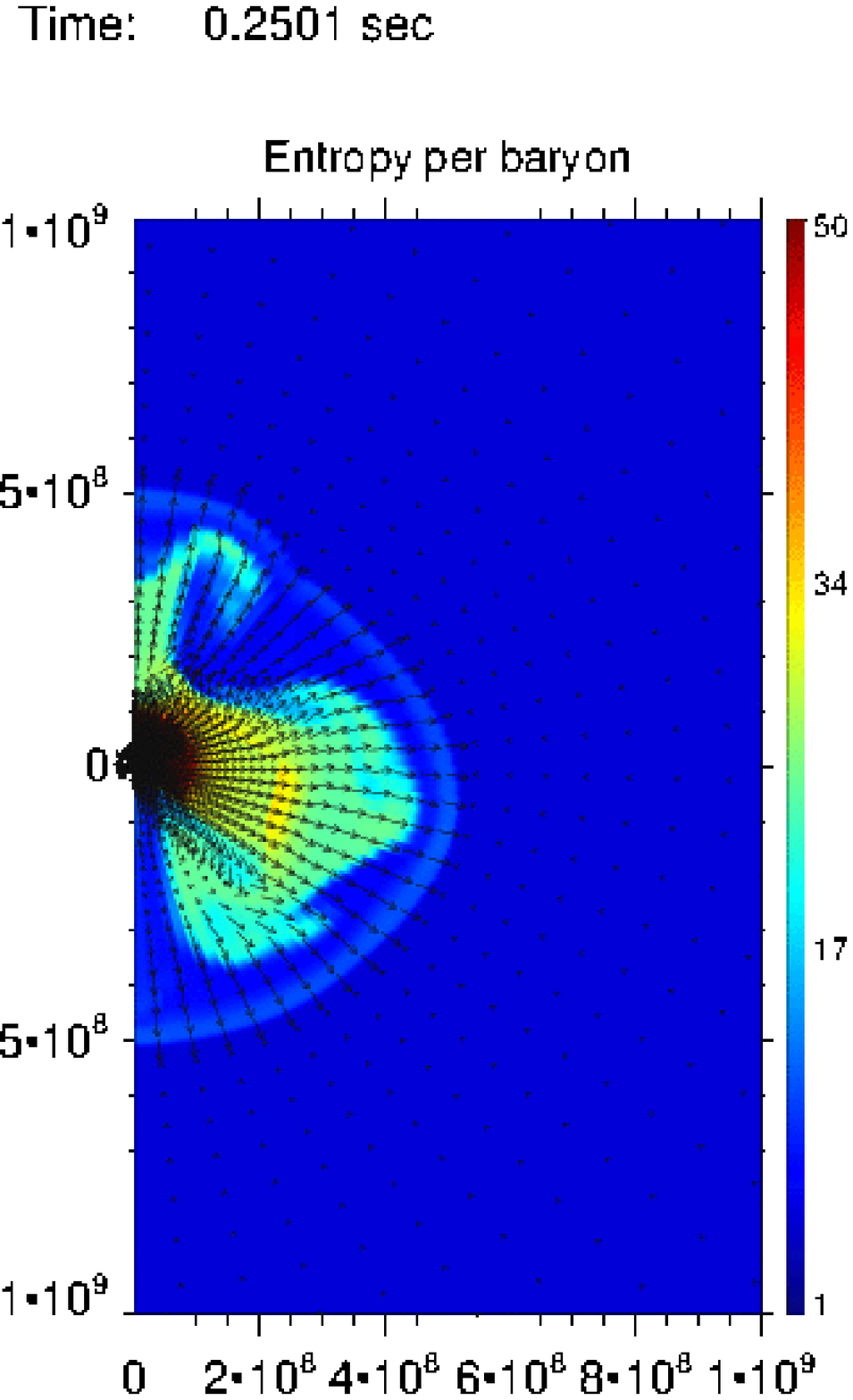}
\plotone{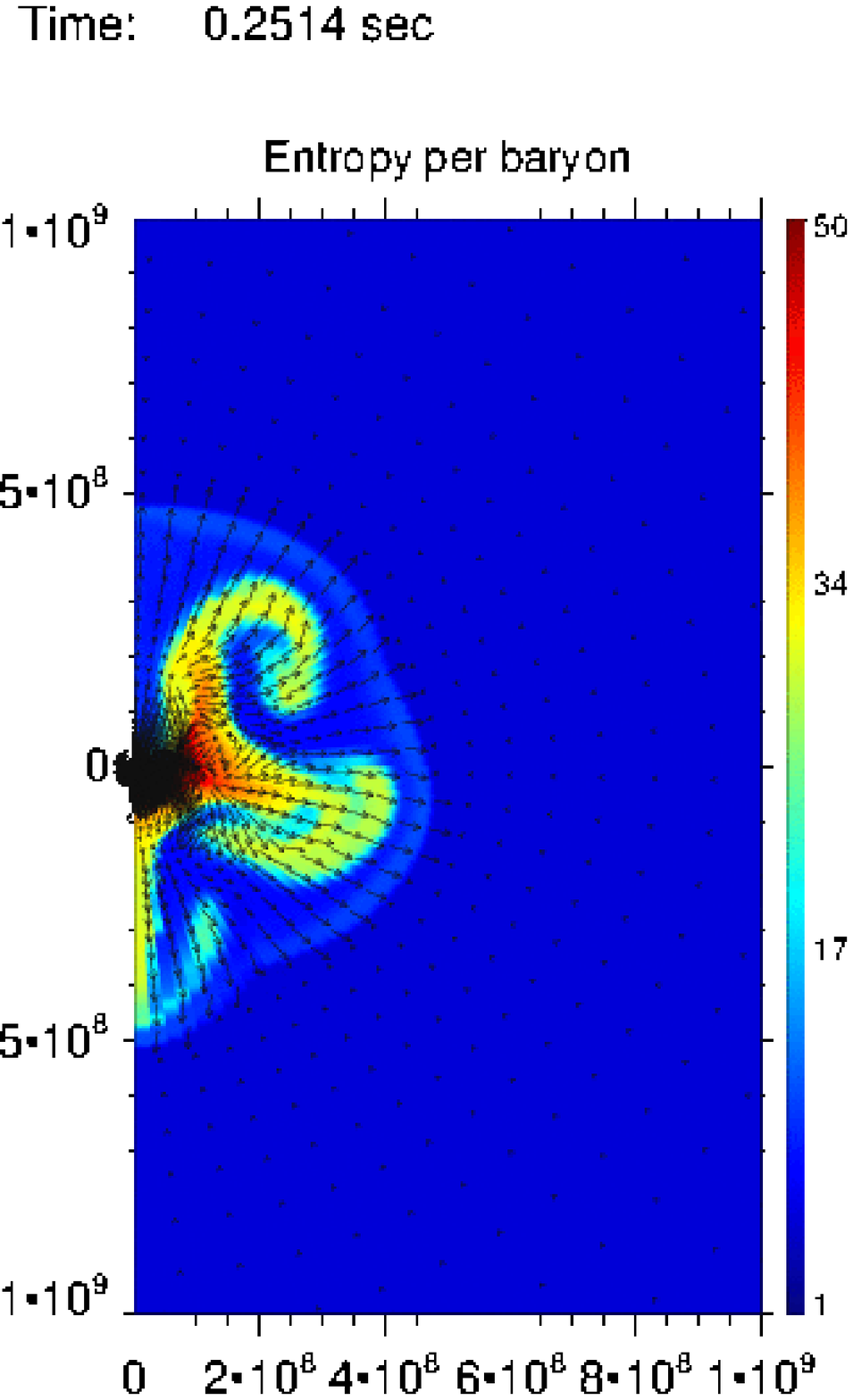}
\plotone{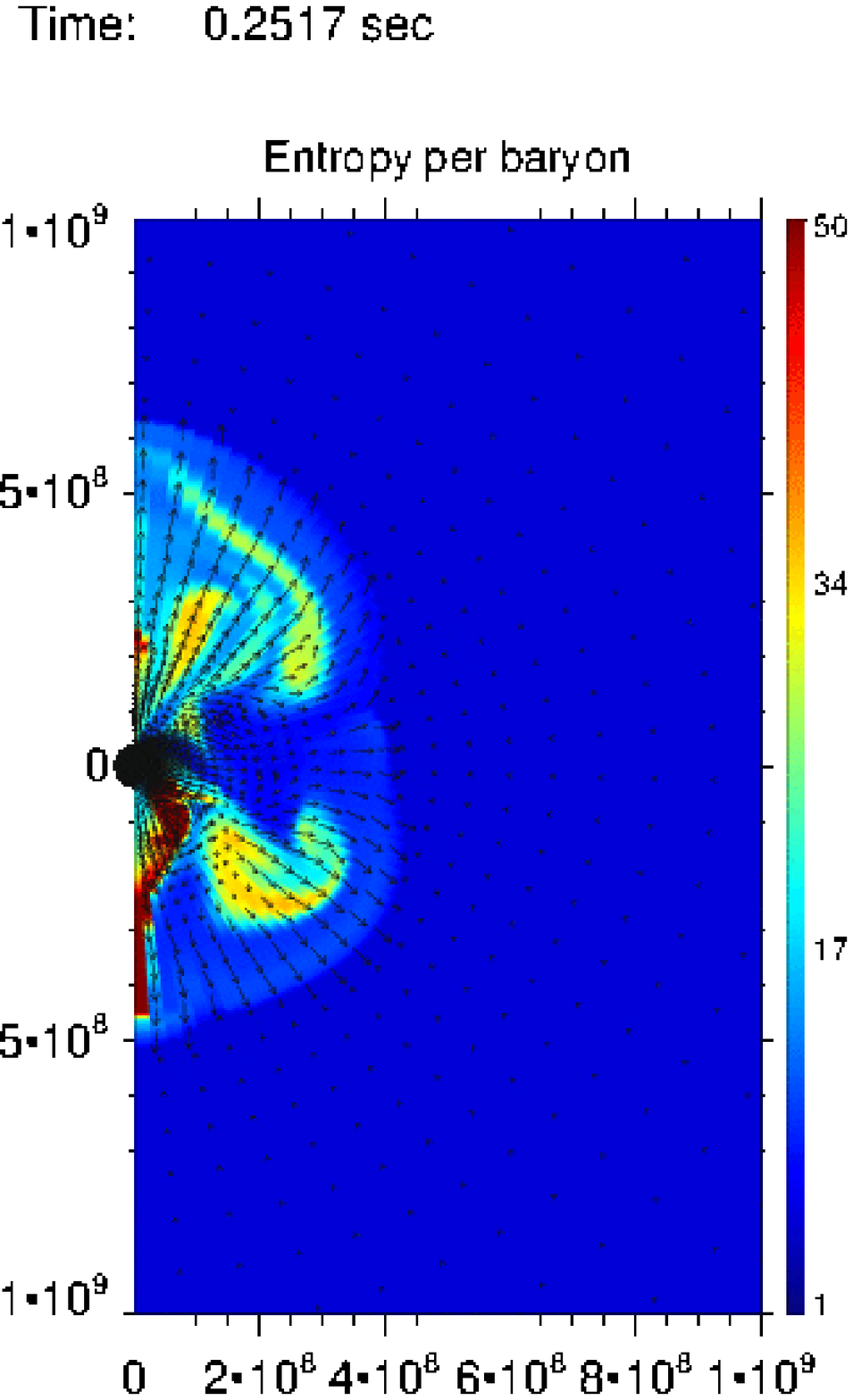}
\plotone{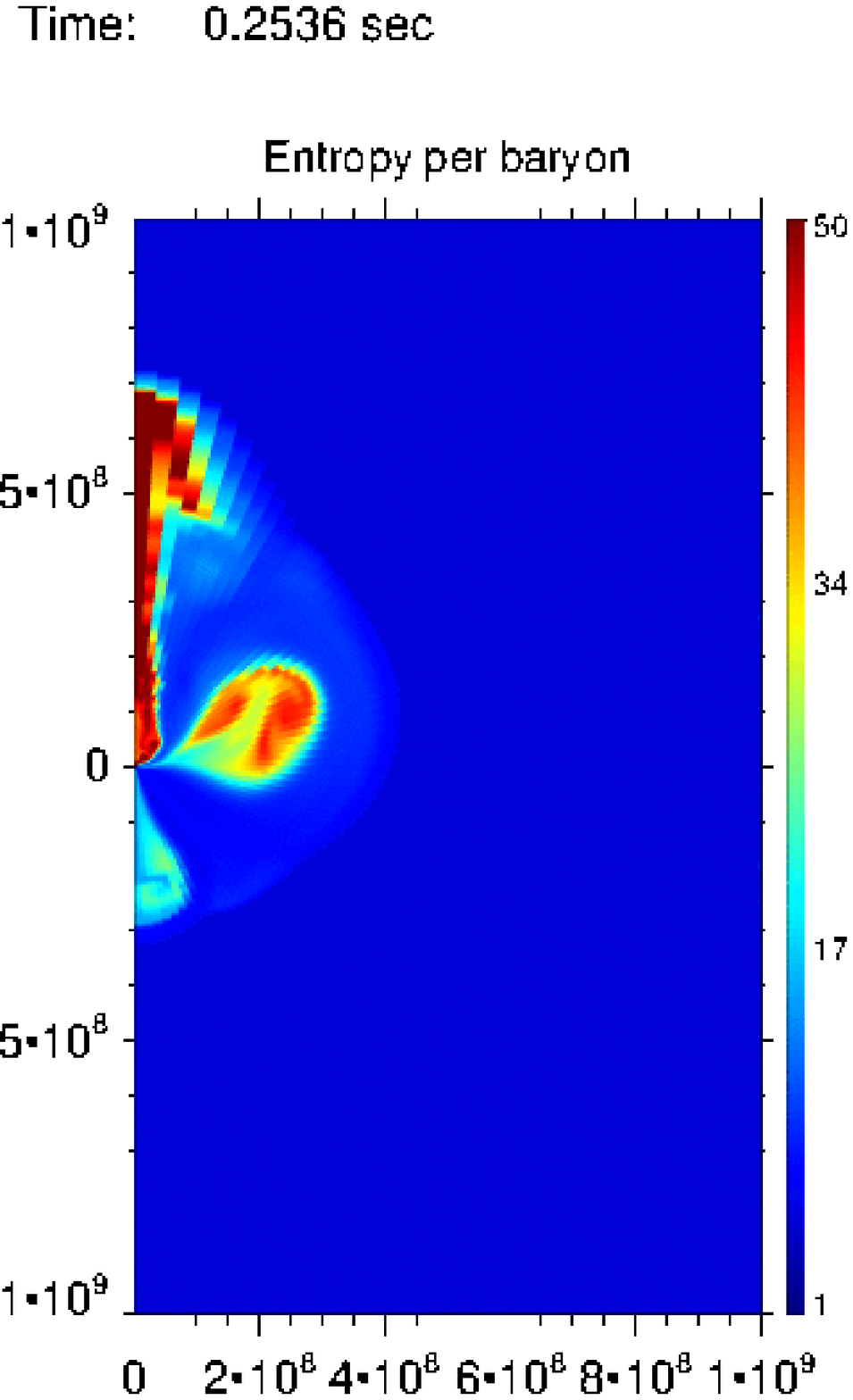}
\plotone{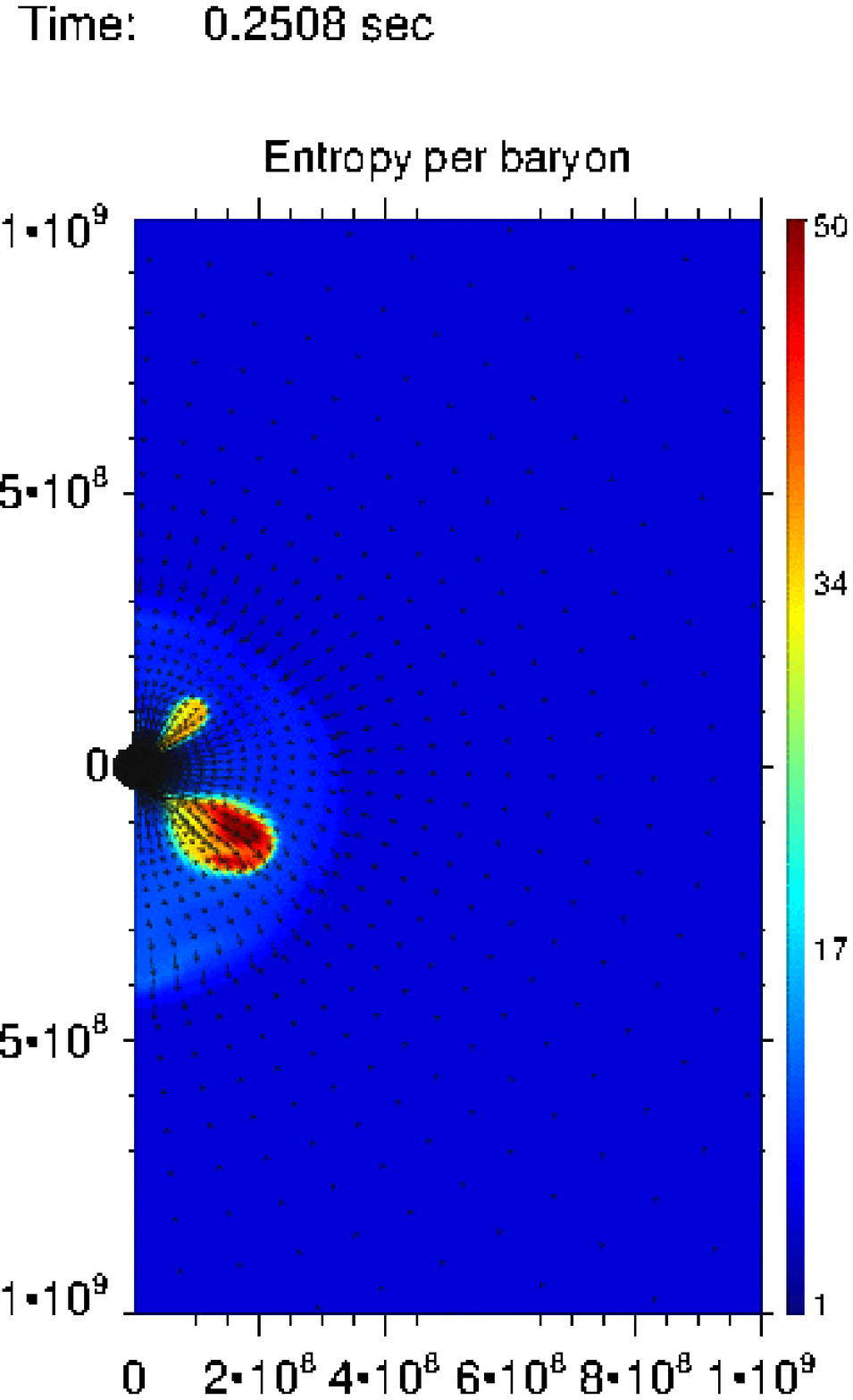}
\caption{Post-shock-relaunch distributions of entropy per baryon in the meridian section for all the 2D models 
at $t_{\rm exp} \sim 0.25 \s$.
\label{fig:shock_shape_2D}}
\epsscale{0.3}
\end{figure*}


\begin{figure}[ht]
 \epsscale{1.0}
 \plotone
{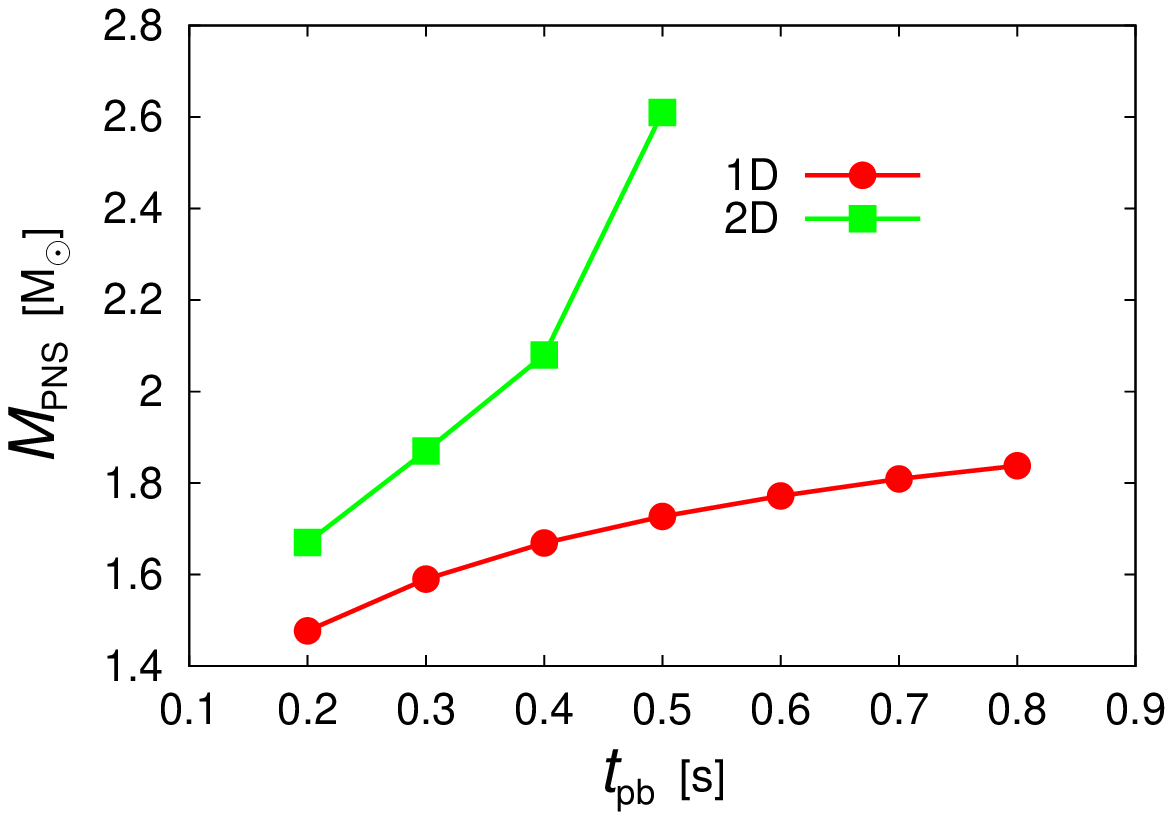}
\caption{Comparison of the masses of proto-neutron stars at the end of computations between the 1D and 2D models.}
\label{fig:pnsmass}
 \epsscale{1.0}
\end{figure}

The post-relaunch evolutions are a bet different between the models with the earlier
($t_{pb}=200$ and 300ms) and later ($t_{pb}=500$ and 600ms) relaunch. 
As shown in Fig.~\ref{fig:shock_shape_2D}, 
SASI is always dominated by the $\ell=1,2$ modes, making the shock front rather prolate 
generically with a marked equatorial asymmetry. 
In the models with the earlier shock revival is nearly isotropic, whereas in the models with 
the later shock relaunch the matter expansion is highly anisotropic
, with large portions of post-shock matter continuing accretions. 
The difference seems to have an origin in the difference of the steady states obtained 
in Step 2. 
In the former the shock radii are large and the post-shock flows are slow. 
As a consequences the matters in the gain region is heated rather homogenously 
in the subsequent evolutions. 
For the latter, on the other hand, the gain region is narrow and the post-shock 
flows are faster, which tends to enhance inhomogenity in the subsequent heating
, leading to the localized expansion. 

The accretion continues
until long after the shock is relaunched. The resultant mass of neutron star is larger in 2D than in 1D as 
pointed out already for the fiducial model and now shown in Fig.~\ref{fig:pnsmass}. 
As the shock revival is delayed
and the critical luminosity gets lower with the mass accretion rate beings smaller, the post-relaunch evolution 
becomes slower as in 1D. This is even more so for the 2D models, for which the critical luminosity is smaller
than for 1D owing to the hydrodynamical instabilities (see Fig.~\ref{fig:fig6}). How these hydrodynamical features 
affect the explosion energy and nickel mass is our primary concern and will be addressed in the next section. 
\subsubsection{Diagnostic explosion energies and masses of $^{56}{\rm Ni}$ in the ejecta}
\label{sec:E-M-2d}

\begin{figure}[ht]
 \epsscale{1.0}
 \plotone{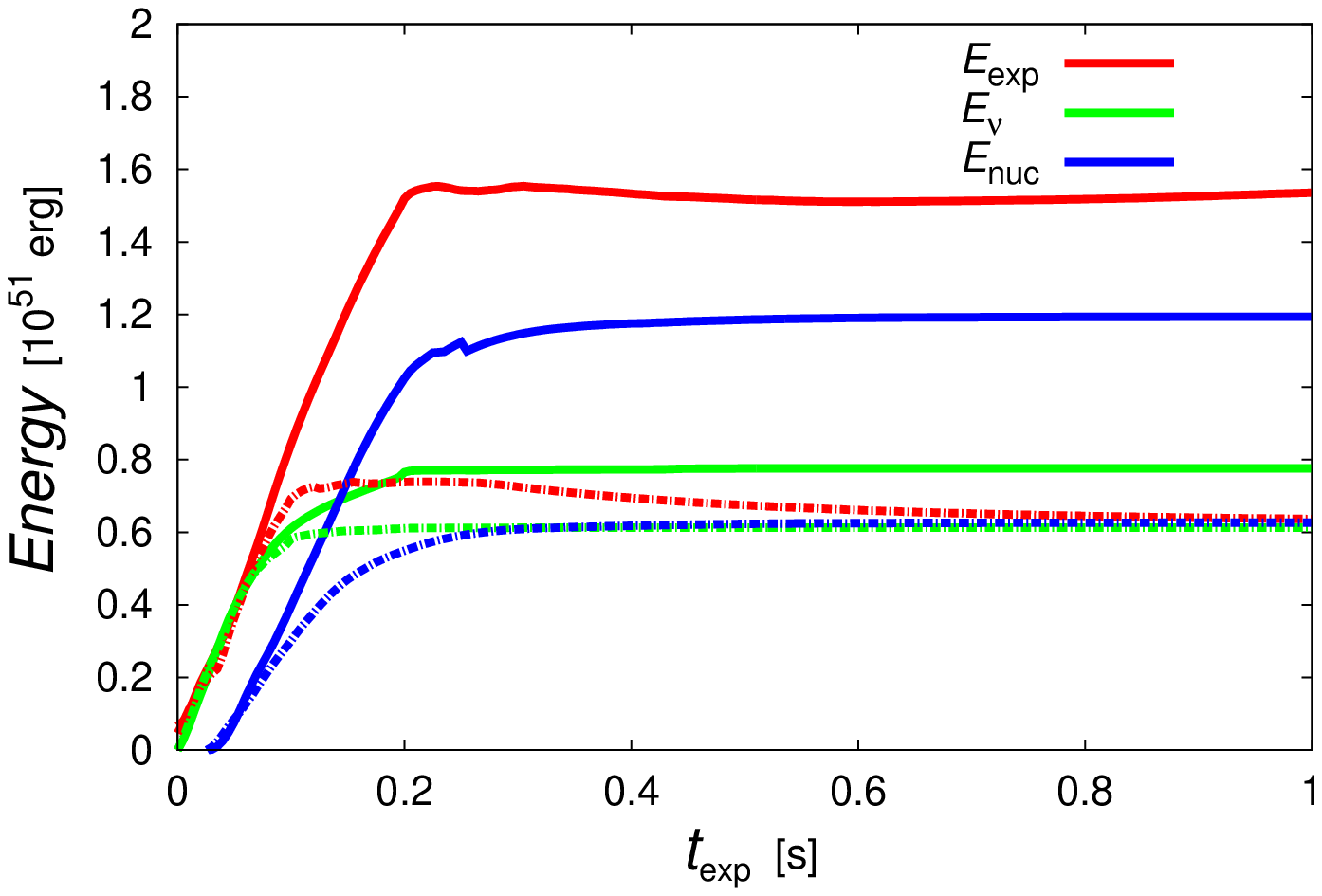}
 \caption{The time evolutions of diagnostic explosion energy for 
the 2D models, in which the shock is relaunched at 
$t_{pb} = 300$ms (solid line) and 
$t_{pb} = 400$ms (solid dotted line).
 }
 \label{fig:Eexp-t400ms-2d}
 \epsscale{1.0}
\end{figure}

\begin{figure}[ht]
 \epsscale{1.2}
 \plotone{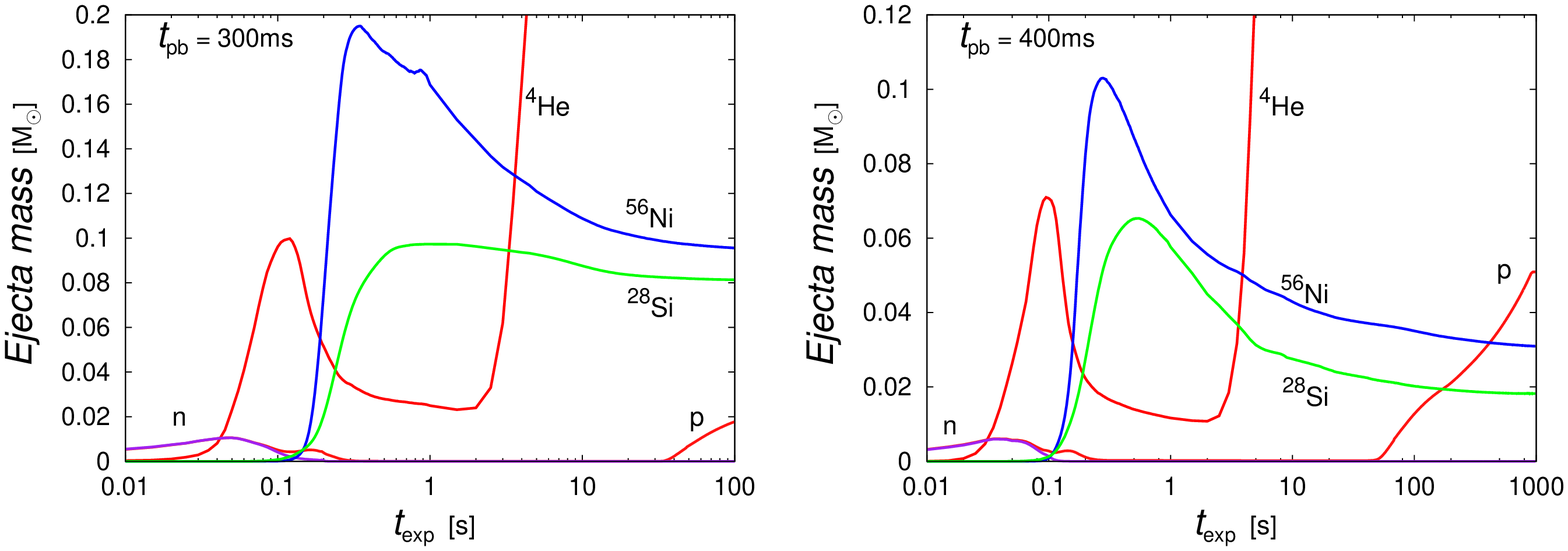}
 \caption{The time evolutions of masses of n, p, \nuc{He}{4}, \nuc{Si}{28}
, and \nuc{Ni}{56} as a function of
$t_{exp}$
for the models with $t_{pb} =$ 300(left panel) and 400ms(right panel).
}
 \label{fig:Mej-t400ms-2d}
 \epsscale{1.2}
\end{figure}

The diagnostic explosion energy is shown in Figure \ref{fig:Eexp-t400ms-2d} as a function of 
$t_{exp}$ for the 2D fiducial model, in which the stalled shock wave is relaunched at
$t_{pb} =$ 300 together with the model with $t_{pb}=400$ms, the counter part to the 1D fiducial model. 
Also presented
are the individual contributions from the neutrino heating and nuclear reactions. As in the 1D fiducial model 
(see Fig.~\ref{fig:fig10}), the neutrino heating is effectively closed at
$t_{exp} \sim 100$ms.
By this time the shock front has reached the location of $r \gtrsim 2,000\km$, which is far enough from the 
proto-neutron star for the neutrino flux to become negligibly small. After the freeze-out of neutrino heating, 
the diagnostic explosion energy increases via the nuclear reactions such as the recombination of \nuc{He}{4} to 
heavier nuclei in the early phase and the Si and O burnings later on as shown in Fig.~\ref{fig:Mej-t400ms-2d},
in which the masses of n, p, \nuc{He}{4}, and $\alpha$ particles, $^{28}{\rm Si}$ and $^{56}{\rm Ni}$ are shown as a function of time. 

The diagnostic explosion energies become almost constant at
$t_{exp}\sim 600$ms in these models and the mass of $^{56}$Ni reach
their maximum values around $t_{exp}\sim 300$ms.
These features are essentially the same as what we observed for the 1D counterpart 
(Figs. \ref{fig:fig8} and \ref{fig:fig10}). 
However, big difference appear after $t_{exp}=300$ms in the masses of heavy elements 
between 1D and 2D: significant fall-backs occur in 2D as can be seen in Fig.19. 
Incidentally, the ratio of the kinetic to internal energies in the ejecta is 
$\frac{E_{kin}}{E_{int}} \sim 4$ at $t_{exp}=1000$sec for all the exploding models. 


\begin{figure}[ht]
 \epsscale{1.0}
 \plotone{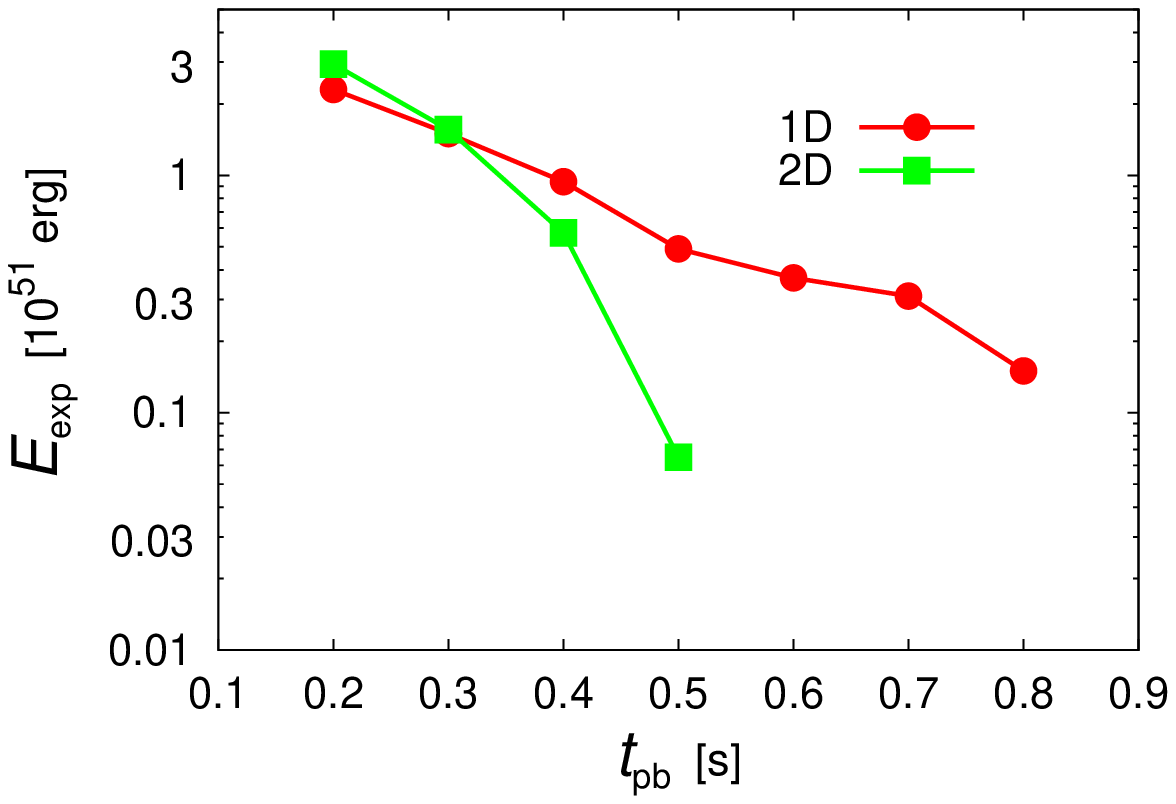}
 \caption{
  Comparison of the explosion energies between the 1D and 2D models.
}
 \label{fig:Eexp-1d2d}
 \epsscale{1.0}
\end{figure}

Figure~\ref{fig:Eexp-1d2d} shows the explosion energies for all the 2D models
in comparison with those for the 1D models. 
It is a general trends that the explosion energy is a monotonically decreasing 
function of the shock-relaunch time although the gradient is much steeper in 2D. 
It is also interesting that the explosion for a given shock-relaunch time is similar 
between 1D and 2D provided the shock revival is early enough to give 
an explosion energy of $10^{51}$erg. 
This may imply that it is the mass accretion rate that primarily determines 
the canonical explosion energy. 
It should be noted, however, that the critical 
luminosity for a give mass accretion rate is smaller in 2D than in 1D (see Fig.~\ref{fig:fig6}).
For a given neutrino luminosity, the explosion energy is hence larger in 2D 
except for very weak explosion by the late shock-relaunch. 
This is the advantage
of non-spherical hydrodynamics that is commonly mentioned in the literature. 
For the model with $t_{pb}=500$ms, the explosion energy is much smaller than 
for the 1D counter part. No explosion obtains in the model with 
$t_{pb}=600$ms owing to the severe fall-back. 
\begin{figure}[ht]
 \epsscale{1.3}
 \plotone{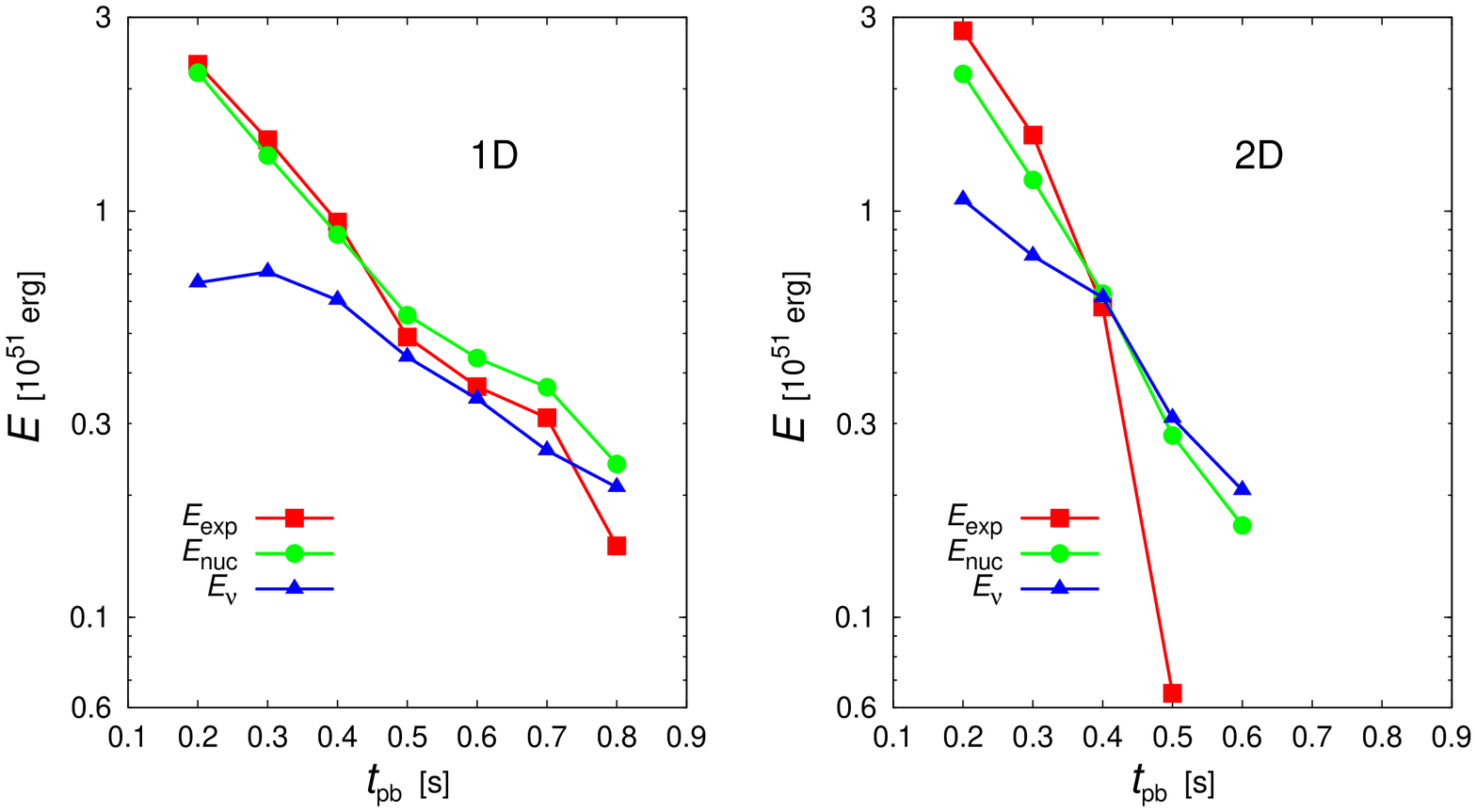}
 \caption{
  The individual contributions of neutrino heating and nuclear reactions to the explosion energy
  for all the 1D and 2D models.
 }
 \label{fig:E-2d}
 \epsscale{1.3}
\end{figure}

In Fig.~\ref{fig:E-2d} we present the individual contributions to the explosion energy from 
the neutrino heating and nuclear reactions 
for both 1D and 2D models, which should be compared with Fig.~\ref{fig:fig12}. 
It is evident that both contributions from the neutrino heating and the nuclear reactions 
drop much more quickly in 2D than in 1D as the shock revival is delayed; 
the contribution of the nuclear reactions decreases faster than that of the neutrino heating 
and , as a consequence, the former is dominant only for the models with earlier 
shock-relaunches ($t_{pb} \lesssim 300$ms). 
It is also interesting that the decay of the explosion energy is 
accelarated once the contribution of the nuclear reactions ceases to be dominant. 
It hence seems that the energy release by the nuclear reactions is an important 
ingrediant for robust explosions. 
The reduction of the contribution from the nuclear reactions is directly related with 
the decrease of the mass elements
in the ejecta that attain high peak temperatures, which we have pointed out already for the fiducial model
(see Fig.~\ref{fig:maxtemp_2d})
, as well as with the fall back in the 2D models. 
These reductions are slightly compensated for by the reduction 
in the (negative) contribution from the gravitational energy of accreting 
matter that is swallowed in the ejecta. 
This is another 
consequence of the fact that the accretion continues in 2D even 
after the shock is relaunched and not all the accreting matter 
is added to the ejecta.


\begin{figure}[ht]
 \epsscale{1.0}
 \plotone{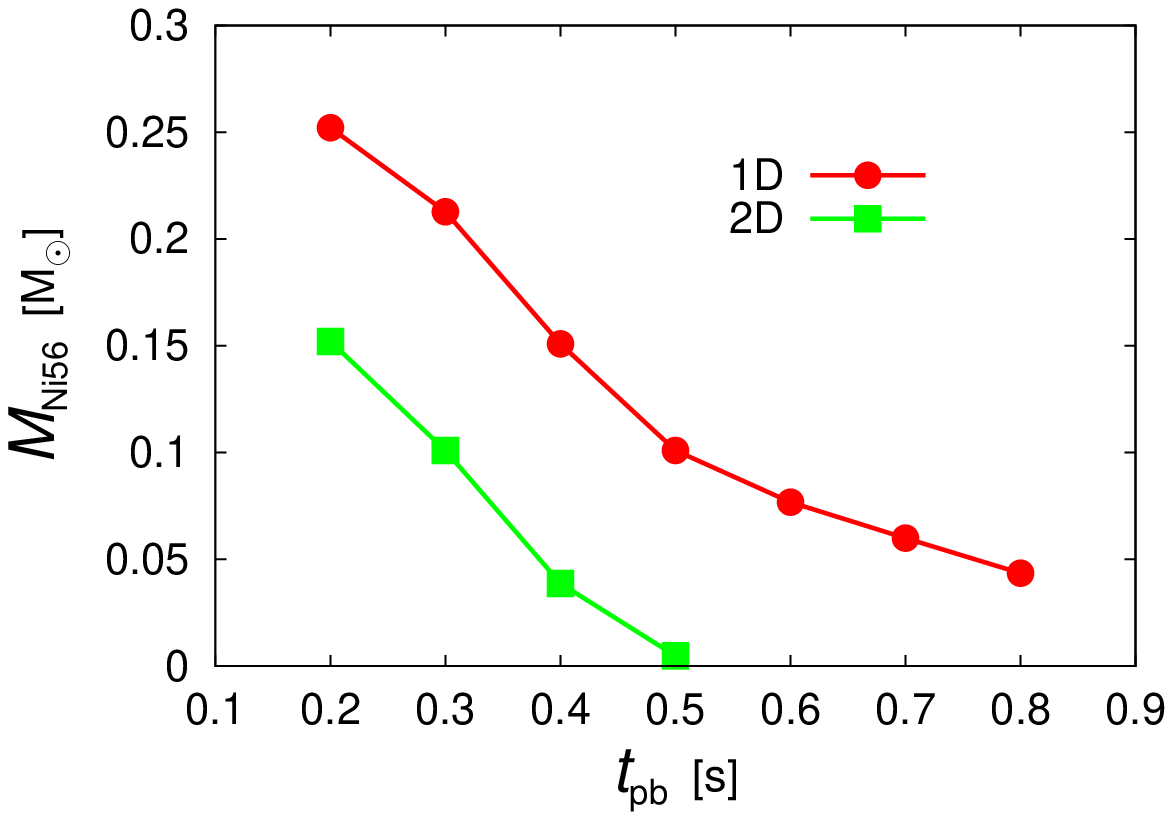}
 \caption{
  Comparison of $^{56}{\rm Ni}$ masses in the ejecta between the 1D and 2D models.
 }
 \label{fig:Mej-1d2d}
 \epsscale{1.0}
\end{figure}

The reduction of the contribution to the explosion energy from the nuclear reactions is also reflected in 
the production of $^{56}{\rm Ni}$, which is demonstrated in Fig.~\ref{fig:Mej-1d2d}. It is apparent that 
the mass of $^{56}{\rm Ni}$ is always smaller in 2D than in 1D as a function of the shock-relaunch time.
It should be reminded that the 1D models tend to over-produce the nickels; as discussed in \S\ref{sec:sys},
if the typical mass of $^{56}{\rm Ni}$ in the ejecta of CCSNe is $M_{\rm Ni} \lesssim 0.1{\rm M}_{\odot}$ as
observations seem to indicate~\citep{2009MNRAS.395.1409S}, the 1D models require that the shock should be relaunched 
later than
 $t_{pb} \sim 500$ms
; this implies a rather weak explosion ($E_{exp}\lesssim 0.5\times 10^{51}{\rm erg}$),
however; no 1D model hence can give both the explosion energy and nickel mass in the appropriate range.
In the 2D models, on the other hand, this problem is much relaxed. Indeed the explosion energy is large 
enough if the shock relaunch occurs earlier than
$t_{pb} \sim 400$ms 
and the nickel is not over-produced
if the shock is revived later than
$t_{pb} \sim 300$ms.
 Although it is entirely a different issue whether and how
the critical luminosity is obtained, this range of shock-relaunch time,
$t_{pb} \sim 300-400$ms,
 may be 
regarded as the appropriate time for shock revival in fact. It is nice that the 2D models have an "allowed" range,
since the hydrodynamics is inevitably non-spherical owing to the hydrodynamical instabilities. 
Whether three-dimensional
hydrodynamics, which is the reality, alters the result for 2D will be an important issue and will be studied in
the future.

Comparison between the abundances in SN ejecta and those in the solar system will possibly lead to the similar allowed range.
\citet{fujimoto11} performed detailed nucleosynthetic calculations for the ejecta of SN explosion from a $15M_\odot$
progenitor~\citep{ww95}.
They were based on simulations of neutrino-driven aspherical explosion, which employed with a hydrodynamic code, that is
similar to the one used in the present study but neglecting the energy generation through nuclear reactions. 
They showed that the explosions with $t_{\rm pb} \sim 200-300{\rm ms}$ give $E_{\rm exp}$ and $M$(\nuc{Ni}{56}) in the allowed range
and that the abundances in the ejecta are similar to those in the solar system.
Detailed nucleosynthesis studies taking into account of the energy generation via nuclear reactions will be interesting, 
since the feedback, which elevates entropy in the ejecta, 
will possibly enhance the amounts of \nuc{Ni}{56} and \nuc{Ti}{44}, which are slightly and highly underproduced 
in \citet{fujimoto11}, respectively.



\section{Summary and discussion}
\label{sec:discussion}

We have investigated by numerical experiments done in 1D and 2D the post-shock-relaunch 
evolutions in the neutrino heating mechanism. Taking into account the fact that the shock 
revival should occur somewhere on the critical line in 
the $L_{\nu,c}-\dot{M}$ diagram 
but exactly where is rather uncertain theoretically at present, we have treated 
the luminosity (or equivalently the mass accretion rate) at the shock relaunch 
as a free parameter that we vary rather arbitrarily. Only the post-bounce phase has been 
computed and we have discarded the central region ($r\lesssim 50{\rm km}$) and replace it with the 
prescribed boundary conditions. We have also neglected the neutrino transport entirely and
employed the light bulb approximation. The shock revival is hence induced by giving
a critical luminosity at the inner boundary by hand.

The critical luminosity itself has been determined also by hydrodynamical simulations,
since not the non-existence of steady state but the onset of hydrodynamical instabilities
dictates the shock revival. The mass accretion rate as a function of time is obtained by
the simulation of gravitational implosion of a massive star envelope. We have adopted
a realistic $15{\rm M}_{\odot}$ progenitor provided by \citet{2007PhR...442..269W}. In these computations,
we have taken into account the nuclear reactions among 28 nuclei that include 14 $\alpha$ 
nuclei as well as their feedback to hydrodynamics consistently. As a result, we have confirmed 
that the critical luminosity is a monotonically decreasing function of the shock-relaunch 
time (or equivalently a monotonically increasing function of the mass accretion rate) and 
that 2D dynamics reduces the critical luminosity compared with 1D dynamics. This is due
to the non-radial hydrodynamical instabilities and the resultant enhancement of neutrino 
heating in the former case.

After re-mapping, we have continued the simulations until long after shock revival. In fact,
for some 1D models we have followed the post-shock-relaunch evolutions up to the shock breakout 
of the stellar surface, confirming that the diagnostic explosion energy has approached the 
asymptotic value much earlier and the shorter computation times for other models are sufficient
indeed. We have employed the same set of input physics as in the simulations for the setup of 
initial conditions. Integrating the source terms in the equation of energy conservation, 
we have evaluated the individual contributions from the neutrino heating and nuclear reactions
to the explosion energy. In so doing, we further divide the latter to the contributions from 
the recombinations in and out of NSE as well as from the nuclear burnings. The axisymmetric 
2D simulations have been performed to elucidate the effect of multi-dimensionality on the outcome. 

What we have found in these model computations are summarized as follows:
\begin{enumerate}
\item
Immediately after shock relaunch the neutrino heating is the dominant source of the explosion
energy but is terminated before long as the shock propagates outwards. Then the nuclear reactions
take its place, with the recombinations of nucleons to $\alpha$ particles under the NSE condition
occurring first. As the temperature decreases, the NSE becomes no longer satisfied. The recombinations
of $\alpha$ particles to heavier elements proceed mainly in this non-NSE circumstance. When the 
temperature lowers further, the nuclear burnings of silicons and oxygens take place in the 
matter that flows into the shock wave. Matter that flows into the shock wave contributes negatively
to the explosion energy, since it is gravitational bound.
\item
The final explosion energy is a monotonically decreasing function of the shock-relaunch time
(or equivalently an increasing function of the mass accretion rate at shock relaunch) irrespective
of the dimensionality of hydrodynamics. 
There is no big difference between 1D and 2D 
for the same mass accretion rate at the shock relaunch as long as 
it occur earlier $t_{pb} \lesssim 300$ms and the explosion is robust. 
The late relaunch in 2D leads to highly anisotropic expansion 
of matter with a large portion of the post-shock matter still 
accreting, which then yield very weak or no explosions. 
This implies that the mass accretion rate 
is the primary factor to determine the canonical 
explosion energy. Since the critical neutrino luminosity
for a given mass accretion rate is lower in 2D than in 1D, the explosion energy for a given
neutrino luminosity is larger 
except for very weak explosion. 
\item
As the shock relaunch is delayed, it takes longer the diagnostic explosion energy reaches the 
asymptotic value. In our 1D fiducial model, in which the stalled shock is revived at
$t_{pb}=400$ms
and we obtain the explosion energy of $E_{exp} \sim10^{51}{\rm erg}$, the diagnostic explosion energy
attains the asymptotic value in
$t_{exp} \sim 600$ms
 whereas it takes $\sim 2{\rm s}$
for the model, in which the shock relaunch is assumed to occur at
$t_{pb}=800$ms
 and the explosion 
energy is as low as $\sim 10^{50}{\rm erg}$. 
The similar trend is also observed 
in the 2D models except for the case with $t_{pb}=600$ms, in which no explosion obtains. 
\item
In 1D the nuclear reactions always overwhelm the neutrino heating. The difference becomes smaller 
as shock relaunch is delayed. To the nuclear reactions the recombinations of nucleons to $\alpha$ 
particles that occur mainly under the NSE condition are the dominant contributor, followed by 
the recombinations of $\alpha$ particles to heavier elements in the non-NSE condition. The nuclear
burnings provide the smallest contribution except for the weakest explosion, which obtains for
the latest shock-relaunch. In that case, the nuclear burnings beat the recombinations in non-NSE.
In 2D, on the other hand, the neutrino heating and nuclear reactions give comparable contributions to 
the explosion energy with 
the latter being larger in stronger explosions and vice versa.
 Note, however,
that the rather small contribution of neutrino heating is deceptive in the sense that it is the ultimate
source of the energy obtained from the nuclear recombinations and that it is crucial to set the stage
for shock revival, which is not accounted for in the diagnostic explosion energy.
\item
In the 1D models nickels are overproduced owing mainly to a larges mass that achieves high peak 
temperatures compared with the ordinary calculation of explosive nucleosynthesis in post-process. 
In fact there is no 1D model that gives the typical values to the explosion energy and nickel mass 
simultaneously. Given observational and theoretical uncertainties, we are not certain how serious a 
problem this is for the moment. One may consider, however, that this is yet another reason to
abandon the 1D neutrino heating model. 
In 2D, on the other hand, this problem is solved, opening up the 
allowed region in the shock-relaunch time around
$t_{pb}\sim 300-400$ms
, which produces the explosion
energy and nickel mass in the appropriate range. 
This happens mainly in 2D because the mass of matter in the 
ejecta that attain high enough peak temperatures is smaller 
and the fall back is significant. 
This is in turn related with the 
fact that the expansion and accretion occur simultaneously in 2D, which is indeed reflected in the mass 
of proto-neutron star, which is larger in 2D at any post-bounce time.
\end{enumerate}

In the present paper we have employed the single $15{\rm M}_{\odot}$ progenitor model, which we think 
is one of the most representative to produce the typical Type IIP CCSN. Very recently \citet{2012arXiv1205.3657U}
reported a possible stochastic nature in the outcome of the shock revival in the neutrino heating mechanism
based on systematic 1D hydrodynamical simulations. Although the stochasticity is less remarkable in 
the low mass end, it is hence mandatory to extend the current work to other progenitors and see how
generic our findings obtained in this paper are \citep{yamamotoetal2012p}. 3D models are also the top priority 
in the future work, since we know that 3D SASI is qualitatively different from 2D SASI we have studied
in this paper \citep{2008ApJ...678.1207I}. It should be also recalled that \citet{2010ApJ...720..694N} claimed that shock revival
is even easier in 3D than in 2D although controversies are still continuing \citep{2012ApJ...755..138H}. 
If the critical luminosity is 
much lower in 3D than in 2D, the yield of $^{56}$Ni may be 
reduced further in 3D. The complex flow pattern also have some 
influences on the nickel yield. 
We are particularly
concerned about how the allowed region in the shock-relaunch time that is opened in 2D is modified in 3D. 
The relative importance of the nuclear reactions for the explosion energy compared with the neutrino heating 
is the highest in 1D. We are certainly interested in what about 3D. 

One of the greatest uncertainties in the present study is the effect of the inner boundary condition that is 
imposed by hand. 
The artificial treatment employed in this paper results in 
the total mass injection from the inner boundary of about 
$7 \times 10^{-3}{\rm M}_{\odot}$ in the 1D fiducial model, 
which contributes to the explosion energies and the \nuc{Ni}{56} masses by 2-3\%. 
Although this may be a slight underestimate \citep{arcones2007}, we believe that 
better treatments will not change the 
conclusion of the paper qualitatively. 
The eventual answer should come from fully consistent simulations of the entire core, though. 
It is also true that the simple light bulb approximation adopted in this paper does not 
accurately account for accretion luminosities, in particular their correlations with temporally 
varing accretion rates as well as the differences between 1D and 2D. 
Hence the appropriate treatment of the neutrino transport, 
which is neglected completely in this paper, will be critically important. 
These caveats notwithstanding we believe that the results obtained in this paper
are useful to understand the post-shock-relaunch evolution in the neutrino heating mechanism, particularly
how the diagnostic explosion energy approaches the final value. One of the goals of our project is to
seek, probably after more systematic investigations suggested above, the way to estimate the explosion 
energy from the early stage of post-shock-revival evolution, since realistic simulations may not be
affordable for a few seconds after shock relaunch.

\acknowledgments{
This work is partly supported by Grant-in-Aid for Scientific Research from the Ministry of
Education, Culture, Sports, Science and Technology of Japan (Nos. 19104006, 21540281, 24740165, 24244036, 22540297), and HPCI Strategic Program of Japanese MEXT.
}




\end{document}